\documentstyle[12pt,aaspp4,tighten]{article}

\def\gtaprx {\lower .1ex\hbox{\rlap{\raise .6ex\hbox{\hskip .3ex
             {\ifmmode{\scriptscriptstyle >}\else
                {$\scriptscriptstyle >$}\fi}}}
                \kern -.4ex{\ifmmode{\scriptscriptstyle \sim}\else
                {$\scriptscriptstyle\sim$}\fi}}}
\def\ltaprx {\lower .1ex\hbox{\rlap{\raise .6ex\hbox{\hskip .3ex
             {\ifmmode{\scriptscriptstyle <}\else
                {$\scriptscriptstyle <$}\fi}}}
                \kern -.4ex{\ifmmode{\scriptscriptstyle \sim}\else
                {$\scriptscriptstyle\sim$}\fi}}}
\def\etal {et al.~}
\def\littleprime{\ifmmode{\scriptscriptstyle \prime }
    \else{\hbox{$\scriptscriptstyle \prime$ }}\fi}
\def\littless{\ifmmode{\scriptscriptstyle s }
    \else{\hbox{$\scriptscriptstyle s $ }}\fi}
\def\littlemm{\ifmmode{\scriptscriptstyle m }
    \else{\hbox{$\scriptscriptstyle m $ }}\fi}
\def\littlehh{\ifmmode{\scriptscriptstyle h }
    \else{\hbox{$\scriptscriptstyle h $ }}\fi}
\def\littlecirc{\ifmmode{\scriptscriptstyle \circ }
    \else{\hbox{$\scriptscriptstyle \circ $ }}\fi}
\def\rasec{\raise .9ex \hbox{\littless}}
\def\arcsec{\raise .9ex
            \hbox{\littleprime\hskip-3pt\littleprime\hskip-3pt}}
\def\ramin{\raise .9ex \hbox{\littlemm}}
\def\arcmin{\raise .9ex \hbox{\littleprime}}
\def\hrs{\raise .9ex \hbox{\littlehh}}
\def\deg{\hbox{$^\circ$}}
\def\degree{\raise .9ex \hbox{\littlecirc}}
\def\magpoint{\hbox to 2pt{}\rlap{\hskip -.5ex \arcmm}.\hbox to 2pt{}}
\def\arcsspoint{\hbox to 1pt{}\rlap{\arcss}.\hbox to 2pt{}}
\def\arcsecpoint{\hbox to 1pt{}\rlap{\arcsec}.\hbox to 2pt{}}
\def\arcminpoint{\hbox to 1pt{}\rlap{\arcmin}.\hbox to 2pt{}}
\def\degreepoint{\hbox to 1pt{}\rlap{\degree}.\hbox to 2pt{}}
\received{2002 March 14}
\begin{document}

\title{The Radio Properties of Composite LINER/H~{\sc ii} Galaxies}

\author{Mercedes E. Filho}
\affil{Kapteyn Astronomical Institute, P.O.~Box 800, NL--9700 AV Groningen, The~Netherlands}

\author{Peter D. Barthel}
\affil{Kapteyn Astronomical Institute, P.O.~Box 800, NL--9700 AV Groningen, The~Netherlands}

\author{Luis C. Ho}
\affil{The Observatories of the Carnegie Institution of Washington, 813 Santa Barbara Street, Pasadena, CA 91101, USA}

\begin{abstract}

Arcsec-resolution  VLA  observations  --  newly obtained  as  well  as
published -- of  40 nearby galaxies are discussed,  completing a study
of  the  radio properties  of  a  magnitude-limited  sample of  nearby
galaxies of  the composite LINER/H~{\sc ii} type.   Our results reveal
an overall detection rate of  at least 25\% AGN candidates among these
composite sources.  The general properties of these AGN candidates, as
compared  to non-AGN composite  sources and  H~{\sc ii}  galaxies, are
discussed.

\end{abstract} 

 
\section{Introduction}

As demonstrated by the extensive studies of Ho, Filippenko, \& Sargent
(1995, 1997a, 1997b), 13\% --  a substantial fraction -- of all nearby
galaxies   is   of  the   ``transition-LINER''   type;  they   display
emission-line  ratios  intermediate  between LINER  galaxies  (Heckman
1980) and  H~{\sc ii}  nuclei.  These  objects are  most  likely truly
composite in nature,  combining weak accretion-driven nuclear emission
with  circumnuclear star-formation (Ho,  Filippenko, \&  Sargent 1993;
see also Ho 1996).  This picture is supported by radio imaging surveys
indicating that  many LINERs  and composite LINER/H~{\sc  ii} galaxies
exhibit  weak compact  nuclear  radio emission,  in  addition to  more
diffuse extended emission (e.g., Filho, Barthel, \& Ho 2000, hereafter
Paper  I; Falcke  \etal 2000;  Nagar \etal  2000, 2002).   Whereas the
latter is most  likely of stellar and interstellar  origin, the former
is  connected   to  nuclear  accretion   processes  and  as   such  is
qualitatively   comparable  to   classical  quasar   emission.   While
dynamical evidence for  black holes is now being  found in many nearby
galaxies  (e.g.,  Richstone  \etal  1998),  the  presence  of  a  high
brightness  temperature  radio core  or  hard  X-ray  emission is  the
clearest evidence  that the black  hole is actually accreting  mass at
the  current epoch.  In  turn, knowing  the demographics  of accreting
black holes, one  can attempt to ascertain the  factors that determine
the incidence and properties of the accretion process.

In  Paper  I,  we  described   our  Very  Large  Array  (VLA)  C-array
observations of a first set  of 25 composite LINER/H~{\sc ii} galaxies
from  the Palomar  sample (Ho  et  al.  1995).   The radio  morphology
combined  with the  radio flux  and spectral  index data  was  used to
isolate AGN candidates  in the composite source sample.   It was found
that roughly 20\% of the  sample objects most likely contain weak AGN.
In a  similar fashion  as Falcke \etal  (2000) and Nagar  \etal (2000,
2002), subsequent milliarcsec  resolution observations have shown that
indeed  these objects  harbour compact,  high  brightness temperature,
flat or inverted spectrum radio cores (Filho, Nagar, \& Barthel 2002),
confirming the  AGN scenario.   In this Paper~II  we apply  the arcsec
resolution  radio  imaging technique  to  the  remaining 40  composite
sources, thereby  completing the analysis of composite  objects in the
full Palomar bright galaxy sample.

In  the following  discussion, we  adopt a  Hubble constant  $H_{0}$ =
75~km  s$^{-1}$  Mpc$^{-1}$  and  define  the  radio  spectral  index,
$\alpha$, such that $F_{\nu} \propto \nu ^{-\alpha}$.

\section{Sample Selection}

All of  the composite LINER/H~{\sc  ii} galaxies presented  in Paper~I
and   in   the   present   paper   were  taken   from   the   original
magnitude-limited Palomar  survey of 486 bright  northern galaxies (Ho
\etal  1995), following  the  classification criteria  outlined in  Ho
\etal (1997a).  The Palomar  sample contains 65 such composite sources
(Ho \etal  1997a), 25  of which  were observed in  1997 and  had their
results  published  in Paper~I.   In  this  Paper~II,  we discuss  VLA
archival data for one source,  published data for 28 more objects, and
new VLA  observations for  the remaining 11.   Properties of  these 40
composite LINER/H~{\sc ii} sample galaxies are given in Table~1.

\medskip 
\centerline{\bf -- TABLE~1: Sample Sources --}

\section {Observations and Data Reduction}

The selection of the composite sources to be observed with the VLA was
based  on  the  absence  of  published,  sensitive  $\ltaprx$5\arcsec~
resolution observations and the  absence of appropriate archival data.
Of these, we choose the  11 sources that fell within the observational
time slot  allocated to  us by  the VLA.  As  will be  dicussed below,
however, the new observations plus  the existing data do permit a more
or less homogeneous assessment of the arcsec-scale radio properties of
the full sample of 65 composite sources.

Similar to the objects in  Paper~I, the 11 galaxies were observed with
the X-band (8.4\,GHz) system of the  VLA, on 2000 March 26.  The array
was  in   its  C-configuration,  yielding  a   typical  resolution  of
2\arcsecpoint5.  Two  bands of 50\,MHz each were  combined, yielding a
total  bandwidth  of  100\,MHz.   Two snapshot  observations  of  6--7
minutes  each, at different  hour angles,  were interspersed  with the
observations of appropriate phase calibrators.  Phase calibrators were
choosen  from the  online  VLA calibrator  manual  to have  positional
accuracies of  the order  of 1~mas.  The  occurence of a  power glitch
during the last hour of our  observing run has resulted in the loss of
half of the data for IC\,356, IC\,1727 and NGC\,1055.

The  flux  densities  were   bootstrapped  by  observing  the  primary
calibrator  3C\,48 (0137+331).  The  adopted 8.4\,GHz  flux densities,
after applying the appropriate baseline contraints as specified in the
online VLA calibrator  manual, were 3.23\,Jy and 3.22\,Jy,  at IF1 and
IF2, respectively.   The calibration  uncertainty is dominated  by the
uncertainty in  the absolute flux  density of the  primary calibrator,
which is conservatively estimated to be 5\%.

Reduction of the data was  performed using standard NRAO AIPS (version
15OCT99) image processing routines.   After minor flagging of the data
and calibration,  the visibilities were Fourier  transformed using the
AIPS task IMAGR.   ``Clean boxes'' were interactively set  in order to
remove  the   sidelobes  from   the  ``dirty  maps,''   following  the
deconvolution CLEAN algorithm (Clark 1980).  The ``cleaning'' task was
halted when the clean flux density reached the noise level as measured
in the  dirty maps.  Full-resolution images,  having synthesized beams
of roughly  2\arcsecpoint5, as well  as tapered maps, having  beams of
about 5\arcsec--7\arcsec~were made.  This proved useful in cases where
weak  low surface  brightness emission  was present.   For comparison,
1\arcsec\ corresponds to  30--300 pc for the large  range of distances
(D=7--70~Mpc) involved.  To within a  factor of two, most images reach
the theoretical noise level of $\sim$0.04~mJy/beam (Perley, Schwab, \&
Bridle 1989).   Self-calibration was not performed  given the weakness
of the radio sources.

1995  VLA  A-array, 8.4\,GHz  archive  observations  of NGC\,660  were
re-reduced  using AIPS as  specified above,  from the  calibrated data
files kindly provided by Dr.   Jim Ulvestad. Tapered as well as a full
resolution image was made.

For  the remaining  28 composite  sources, we  have  gathered relevant
published radio  data, at  cm-wavelengths and resolutions  between 0.2
and 5~arcsec.  These are  summarized in Table~5. 

\section{Results}

\subsection{Radio images}

Overall, 10 of the 11  newly observed composite sources were detected.
Only IC\,1727 remains undetected and we establish a conservative upper
limit  of  0.5~mJy  (5$\sigma$)   to  any  radio  emission  (see  also
Section~4.2).   Table~2 provides  the map  parameters of  the detected
sources,  including NGC\,660.  The  map rms  has been  estimated using
AIPS  task IMSTAT,  in a  source-free region.   Images of  the sources
appear  in Figures  1 --  3, in  the Appendix.   We present  the radio
contour  maps (typical beamsize  2\arcsecpoint5) overlayed  on optical
images  taken from  the Digitized  Sky Survey  (DSS),  with contouring
according  to CLEV $\times$  ($-$3,3,6,12,24,48,96) mJy/beam  (CLEV is
the map noise  level; see Table~2).  For the  full resolution image of
NGC\,660 (0\arcsecpoint21~resolution) and for NGC\,1055 and NGC\,7331,
having substantially  more complex  radio morphologies than  the other
sources, we  present radio  grey scale images,  with the  scale levels
given above each image.

\medskip 
\centerline{\bf -- TABLE~2: Radio map parameters --}

Given that  good phase stability was  obtained as judged  from the VLA
phase calibrators,  the astrometric accuracy of  the overlay procedure
must be  dominated by  the DSS  accuracy, which is  known to  be about
0.6\arcsec~(V\'eron-Cetty \&  V\'eron 1996).  This is born  out by the
images of  several unresolved and slightly  resolved nuclei (NGC\,410,
NGC\,521, NGC\,524) which appear accurate to within a few tenths of an
arcsec.

We have  also detected unrelated  background radio sources  in several
fields:  their  number is  entirely  consistent  with the  statistical
probability  of finding such  sources in  a field  of a  given primary
beamsize and sensitivity.  Due to their large distances from the field
center,  primary beam  corrections were  applied to  infer  their flux
densities.  NED  (NASA/IPAC Extragalactic Database)  and the Cambridge
APM  (Automatic  Plate  Measuring)  facility  have  proven  useful  in
identifying some of the background  sources.  In Table~3 we list basic
properties of  these sources, where  we have designated them  by their
sky orientation  with respect  to the field  center, for  the relevant
galaxy.  These are discussed below.

\medskip \centerline{\bf -- TABLE~3: Field source parameters --}

\subsection {Comments on Individual Galaxies}

In this section  we present short discussions on  the sample galaxies,
starting with  the newly observed galaxies  (plus background sources).
We generally take  into consideration data from the  VLA 1.4\,GHz NVSS
and  FIRST surveys  (45\arcsec~and  5\arcsec~resolution, respectively;
Condon  \etal  1998a; Becker,  White,  \&  Helfand  1995), Green  Bank
4.85\,GHz survey (3\arcminpoint5~resolution; Becker, White, \& Edwards
1991) as well as other available radio data.

{\it  IC\,356:}  NVSS  did  not  detect this  source.   The  1.4\,GHz,
54\arcsec~resolution  observations  of  Condon  (1987)  did  detect  a
diffuse 29.5~mJy  source.  We detect a  slightly resolved, $\sim$1~mJy
core, coincident with the optical nucleus of the galaxy.

There is also  another object (IC356NE), having about  6~mJy, at about
3\arcminpoint5  from the  target source.   We  have not  been able  to
identify this source  using NED, which indicates that  the two nearest
objects   are   87GB\,040255.7+694248   and  IRAS\,04033+6942,   about
1\arcminpoint3 away.  It may  possibly be identified with the 80.1~mJy
NVSS  source at a  distance of  5\arcsec~from  our radio  position for
IC356NE. There are no APM catalogue data for this field.

{\it IC\,1727:} There is no  NVSS detection of this source.  1.4\,GHz,
54\arcsec~resolution  observations  (Condon  1987) show  an  uncertain
1.3~mJy detection.   Although our full  resolution map of  this source
suggests the  presence of very  weak low surface  brightness emission,
tapering showed  just noise.  The non-detection could  also be related
to the loss of part of  the data, and thereby loss of sensitivity.  We
establish an upper limit of  0.5~mJy (5$\sigma$) to any radio emission
at the resolution achieved in our C-array observations.

We  do, however, detect  a weak  (3.5~mJy) Southwestern  source, about
3\arcminpoint5 from our target center.  Consultation of NED shows that
the closest  object is the NVSS source  J014701+272105, 5\arcmin away.
The  APM data  show  two  red objects  ($m_{\rm  R}$=19.25 and  19.28,
respectively) at a  distance of about 5\arcsec~on either  side of this
position  -- their  nature with  respect to  the radio  source remains
unclear.

{\it NGC\,410:}  NVSS detected a  6.3~mJy source.  We have  detected a
$\sim$1.7~mJy unresolved core, coincident  with the optical nucleus of
the galaxy.

{\it   NGC\,488:}  NVSS   did  not   detect  this   source.   However,
54\arcsec~resolution  observations at  1.4\,GHz (Condon  1987)  show a
diffuse 7.6~mJy  source --  the radio emission  must have  low surface
brightness.  Both our full resolution and tapered map show only a very
weak,  $\sim$0.3~mJy unresolved  core, at  the optical  center  of the
galaxy.

{\it NGC\,521:} This  source was not detected in  the NVSS.  1.4\,GHz,
0\arcminpoint2~resolution observations of  Hummel \etal (1985) give an
upper  limit of  4~mJy  to any  radio  emission.  We  have detected  a
slightly  resolved $\sim$0.5~mJy core  in the  optical nucleus  of the
galaxy.

The  NGC\,521 field  also shows  two  other sources  to the  Southeast
($\sim$3\arcmin)   and  Northeast   ($\sim$3\arcmin)  which   we  have
designated as NGC\,521SE and NGC\,521NE, respectively.  NGC\,521SE has
been identified  with the 17.6~mJy NVSS source  J012437+014112.  It is
also seen on both the red and blue APM plates as a stellar object with
blue magnitude 21.74.  NGC\,521NE has been identified with the 5.7~mJy
NVSS   source   J012443+014535.    However,   there  is   no   optical
identification on either the red or blue APM plate.

{\it   NGC\,524:}  NVSS   reports  a   3.5~mJy  source.    At  5\,GHz,
5\arcsec~resolution, Wrobel \& Heeschen (1991), detect a 1.4~mJy core.
Our higher resolution observations show a slightly resolved 2~mJy core
in the center  of the optical galaxy.  Comparison  of our measurements
with the NVSS  and Wrobel \& Heeschen (1991)  flux densities indicates
the source to be compact and have a flattish spectrum.

We  have  also  detected  a  $\sim$0.7~mJy  source  to  the  Southwest
(NGC\,524SW), about  2\arcmin from the  galaxy.  As given by  NED, the
closest   object   to   NGC\,524SW   is  the   supernova   SN\,2000cx,
1.4\arcmin~away.   The  nearest  NVSS  source  is  at  a  distance  of
1\arcminpoint8.  There is no  optical identification on either the red
or blue  APM plate, although there  is a noise-like object  on the red
plate with magnitude 19.85, less than 4\arcsec~away.

{\it NGC\,660:}  NVSS detected a  373~mJy source, consistent  with the
54\arcsec~resolution  observations  of   Condon  (1987)  at  the  same
frequency.  184~mJy were detected in the Green Bank survey.  5\arcsec,
1\arcsecpoint5,  and   0\arcsecpoint5  resolution  maps   at  1.4  and
4.85\,GHz (Condon  1980; Condon \etal 1982; Condon  1987; Condon \etal
1990) show  a double structure  in the  Northeast-Southwest direction.
High  resolution, 15\,GHz  observations (Carral,  Turner, \&  Ho 1990)
show  an edge-on  ring or  linear  jet structure  composed of  several
aligned components at a position angle of 71\deg.  The optical nucleus
is located about 3\arcsec~South of the strongest 15\,GHz peak.

The overall  radio morphology of NGC\,660  measures about 6\arcsec~and
appears co-spatial  with the small  optical galaxy.  From  the tapered
($\sim$2\arcsecpoint5~resolution)  image, we  compute  the total  flux
density to be about 60~mJy.   The full resolution 8.4\,GHz map shows a
beautiful  symmetric ring of  radio emission  and what  looks to  be a
radio source in  its center.  We shall designate  this radio component
as the `core' and we  estimate it comprises less than 3~mJy.  Although
individual points of emission can  be identified, we call attention to
two strong  Northeastern components, about  2--2\arcsecpoint5 from the
core;  these have  also been  observed  by Carral  \etal (1990).   For
consistency with the Carral \etal  (1990) maps we have designated them
components C1 and  C2.  Comparison of our and  the Carral \etal (1990)
measurements of  these components shows  that they are  steep spectrum
and/or variable, hence possible supernova remnants.

{\it  NGC\,1055:}  NVSS reports  a  201~mJy  source,  and 85~mJy  were
detected  in the  Green Bank  survey.   1.4\,GHz, 54\arcsec~resolution
observations  (Condon 1987)  show  a 302~mJy  source  oriented in  the
East-West direction.  FIRST  shows an 7~mJy peak cradled  by a disk of
radio emission, totalling 19.3~mJy.  Higher resolution observations at
the same  frequency (Downes  \etal 1986) show  that most of  the radio
emission arises from the galactic disk.

Similar  to the FIRST  image, our  observations show  edge-on ringlike
radio emission over about  2\arcmin, co-spatial with the galactic disk
in  addition  to  a  relatively strong  $\sim$7~mJy  core,  apparently
originating from  the optical nucleus  of the galaxy.  The  total flux
density,  as judged from  tapered maps  ($\sim$7\arcsec~resolution) is
about 40~mJy.   Comparison of our  observations with the  FIRST images
shows the core to be compact and have a flat spectrum.

{\it NGC\,1161:} NVSS detected a 4.9~mJy source.  We detect a slightly
resolved, $\sim$4~mJy core, coincident with the optical nucleus of the
galaxy.  Comparing this with the NVSS  value shows the source to be of
a compact and flat spectrum nature.

We also detect a source  (NGC\,1161S), about 2\arcmin due South of the
target galaxy.  While  neither NED nor APM contain  this object, there
is a 3.9~mJy NVSS source coincident with NGC\,1161S.

{\it  NGC\,7177:} NVSS  reports a  28~mJy source  consistent  with the
1.4\,GHz, 54\arcsec~resolution observations  of Condon (1987).  Hummel
\etal (1987)  establish an upper limit  of 0.8~mJy to  any emission at
1.4\,GHz,  1\arcsecpoint3 resolution.  We  detect a  somewhat resolved
0.8~mJy core, in the center of the optical galaxy; comparison with the
integrated  NVSS flux  density otherwise  indicates  strong resolution
effects.

{\it  NGC\,7331:} NVSS data  imply strong  resolution effects  for the
$\sim$2\arcmin~  North-South extended  radio emission  associated with
NGC\,7331.   The   Green  Bank  survey  measured   an  80~mJy  source.
1.49\,GHz,  54\arcsec~resolution   map  of  Condon   (1987),  shows  a
North-South,  373~mJy  source.   1\arcsecpoint5~ resolution,  1.4  and
5\,GHz maps were obtained by Cowan, Romanishin, \& Branch (1994).  The
maps show a central unresolved  flat spectrum radio source of 0.23 and
0.12~mJy, respectively and a ring of radio emission around the central
source, containing  mainly nonthermal  sources which may  be supernova
remnants.

Similar to the high resolution image  of NGC\,660, we detect a ring of
radio emission,  co-spatial with  the galaxy major  axis.  Due  to the
complexity of the  source and the reduced sensitivity  relative to the
observations of  Cowan \etal  (1994), we were  not able to  detect the
radio core.  As  judged from tapered maps ($\sim$7\arcsec~resolution),
the total 8.4\,GHz flux density is approximately 50~mJy.

We also detect  a strong source (NGC\,7331SW), 4\arcmin  away from the
target center, to  the Southwest.  Since this source  is almost at the
primary beam edge,  our estimate of $\sim$20~mJy should  be taken as a
lower limit.  NED shows that the closest source, 6\arcsec~away, is the
423~mJy radio source TXS\,2234+340  (Douglas \etal 1996).  There is no
NVSS detection, and APM data for this field are as yet lacking.

{\it   NGC\,7742:}   NVSS   detected   a  27~mJy   source.    However,
0\arcsecpoint2~resolution observations of Nagar \etal (2000) establish
an upper limit of 1.1~mJy to  any small scale emission at 15\,GHz.  We
detect  approximately  20~mJy  of  low  surface  brightness  emission,
extending over most of the galaxy disk.

We  also   detect  a  source  to  the   Southwest  (NGC7742SW),  about
1\arcminpoint5 away from  the target.  The nearest source  as given by
NED is  supernova SN\,1993R, 1\arcminpoint7 away and  the closest NVSS
source is  a 4.5~mJy  object, about 2\arcsec~away.   Both the  red and
blue APM plates show there is a nonstellar source ($m_B$ = 20.82 mag),
about 7\arcsec~from the target position.

{\it The  following 28 sample  objects were not  observed by us  -- we
briefly  discuss their  literature  data  as relevant  to,  and in  the
context of, our observations.}

{\it NGC\,2859:} This  source was not detected in  the NVSS.  Reported
upper limits for any radio  emission at moderate resolution are 10 and
0.5~mJy at 1.4\,GHz and  5\,GHz, respectively (Hummel \& Kotanyi 1982;
Fabbiano \etal 1987; Wrobel \& Heeschen 1991).

{\it NGC\,3245:} NVSS  reports a 7.2~mJy source, similar  to the 6~mJy
detected within  FIRST.  Wrobel \& Heeschen (1991)  find an unresolved
3.3~mJy core at 5\,GHz, 5\arcsec~resolution.  Judging from these data,
the source could be flat spectrum and compact.

{\it  NGC 3489:}  NVSS did  not  detect this  source.  Other  1.4\,GHz
observations  at moderate resolution  quote an  upper limit  of 10~mJy
(Hummel \& Kotanyi  1982; Fabbiano \etal 1987) to  any radio emission.
5\arcsec~ resolution, 5\,GHz observations of Wrobel \& Heeschen (1991)
give an upper limit of 0.5~mJy.

{\it  NGC\,3692:}  NVSS  detected  a  8.4~mJy  source.   The  15\,GHz,
4\arcsec~and  0\arcsecpoint2~resolution  observations  of Nagar  \etal
(2000) yield an  upper limit of 1.2~mJy  to any radio  emission -- its
nature must be diffuse and of low surface brightness.

{\it NGC\,3705:}  There is no  NVSS detection of this  source.  Condon
(1987),  1.4\,GHz observations at  54\arcsec~resolution show  a 21~mJy
Northwest-Southeast  elongated   source.   At  the   same  resolution,
10.7\,GHz observations  measure a  10~mJy source (Niklas  \etal 1995).
The  15\,GHz, 4\arcsec~and  0\arcsecpoint2~resolution  observations of
Nagar \etal (2000) give an upper limit of 1.2~mJy to any compact scale
radio emission.

{\it NGC\,3898:}  NVSS reports  a 8.7~mJy source  and FIRST  a 2.4~mJy
source.   The 1.4\,GHz,  54\arcsec~resolution  observations of  Condon
(1987) detect  an uncertain  6.3~mJy source,  spatially confused  by a
strong, nearby, unrelated  core.  The same radio structure  is seen in
the Gioia \etal (1987) map.

{\it NGC\,3917:} NVSS  and FIRST did not detect  this source.  Condon,
Yin,  \&  Burnstein (1987)  show  an  uncertain  2.1~mJy detection  at
1.4\,GHz,    54\arcsec~resolution.     The    15\,GHz,    4\arcsec~and
0\arcsecpoint2 resolution  observations of Nagar \etal  (2000) give an
upper limit of 1.2~mJy to any small-scale radio emission.

{\it NGC\,3953:}  This source  was not detected  in the  NVSS.  Condon
(1987) shows    a    45~mJy    North-South   source    at    1.4\,GHz,
54\arcsec~resolution.   Single-dish  observations  at  5  and  11\,GHz
(Garcia-Barreto \etal  1993) show that this  source is steep-spectrum.
Hummel (1980) establish an upper  limit of 10~mJy to any core emission
at   1.4\,GHz,  while   the   15\,GHz,  4\arcsec~and   0\arcsecpoint2~
resolution observations of  Nagar \etal (2000) give an  upper limit of
1.3~mJy to any radio emission.

{\it    NGC\,3992:}    NVSS    did    not    detect    this    source.
54\arcsec~resolution,  1.4\,GHz  observations  (Condon  1987)  show  a
21.3~mJy  diffuse  source.  High  resolution  observations at  various
frequencies  (Heckman 1980;  Hummel \etal  1985;  Garcia-Barreto \etal
1993) show this  object to harbour a  flat-spectrum, $\sim$2~mJy core.
However,   for  unclear   reasons,  the   4\arcsec~and  0\arcsecpoint2
resolution 15\,GHz observations of  Nagar \etal (2000) give only upper
limit of 1.3~mJy to any radio emission.

{\it NGC\,4125:} There is no NVSS detection of this source.  1.4\,GHz,
18\arcsec~resolution  observations  (Condon  \etal  1998b)  detect  an
uncertain 1.8~mJy source.  Moderate-resolution observations at 1.4 and
5\,GHz (Heckman  1980; Hummel 1980;  Hummel \etal 1983) give  an upper
limit  of 10~mJy  to  any core  emission,  while the  high-resolution,
5\,GHz observations  of Wrobel \& Heeschen (1991)  establish a 0.5~mJy
upper limit.

{\it  NGC\,4145:}   NVSS  did  not  detect   this  source.   1.4\,GHz,
54\arcsec~resolution  observations (Condon  1987) detect  an uncertain
2.8~mJy  source,   while  the  15\,GHz,   4\arcsec~and  0\arcsecpoint2
resolution observations of  Nagar \etal (2000) give an  upper limit of
1.3~mJy to any small-scale radio emission.

{\it NGC\,4150:} This  source was not detected in  the NVSS.  1.4\,GHz
15\arcsec~resolution  observations (Condon  \etal 1998b)  show  a weak
0.8~mJy source.  The high-resolution, 5\,GHz observations of Wrobel \&
Heeschen (1991) establish a 0.5~mJy upper limit to any radio emission.

{\it  NGC\,4192:} NVSS  reports  a 24~mJy  source,  while 33~mJy  were
detected  in the  Green Bank  survey.   1.4\,GHz, 54\arcsec~resolution
observations (Condon 1987) show  a 74~mJy North-South oriented source.
FIRST detected a 16.8~mJy  source, slightly extended to the Southeast,
while 15\,GHz, 4\arcsec~resolution  observations of Nagar \etal (2000)
detect  1.3~mJy  --  the  source  must be  dominated  by  low  surface
brightness emission.  The same authors  give an upper limit of 1.3~mJy
to  any   small-scale  radio  emission   at  15\,GHz  (0\arcsecpoint2
resolution).

{\it NGC\,4216:}  NVSS detected a 7~mJy  source.  1.4\,GHz, 54\arcsec~
resolution  observations (Condon 1987)  detect a  13.4~mJy North-South
source.  However,  this source was  not detected at 10.7\,GHz,  at the
same  resolution (Niklas  \etal  1995).  15\,GHz,  4\arcsec~resolution
observations of  Nagar \etal (2000) detect 2.0~mJy.   The same authors
give an  upper limit of 2.0~mJy  to any small-scale  radio emission at
15\,GHz (0\arcsecpoint2~resolution.)

{\it  NGC\,4220:} NVSS  reports a  4~mJy source.   Wrobel  \& Heeschen
(1991) establish an upper limit of  0.5~mJy to any emission at 5\,GHz,
5\arcsec~resolution, while  the 4\arcsec~and 0\arcsecpoint2~resolution
observations of Nagar  \etal (2000) give an upper  limit of 1.4~mJy to
any radio emission at 15\,GHz.

{\it NGC\,4281:} NVSS  did not detect this source.   There is an upper
limit  of  0.5~mJy  to  any emission  at  5\,GHz,  5\arcsec~resolution
(Wrobel \& Heeschen 1991).

{\it NGC\,4324:} There is no NVSS detection of this source.  Wrobel \&
Heeschen (1991) establish an upper limit of 0.5~mJy to any emission at
5\,GHz, 5\arcsec~resolution.

{\it  NGC\,4350:}  NVSS  did  not  detect  this  source.   At  5\,GHz,
5\arcsec~resolution,  there  is  an  upper  limit of  0.5~mJy  to  any
emission (Wrobel \& Heeschen 1991).

{\it  NGC\,4419:}   NVSS  reports  a  55.4~mJy   source  and  1.4\,GHz
observations at 15\arcsec~resolution (Condon \etal 1990) show a 50~mJy
source.     FIRST   reports    a    39.5~mJy   detection.     15\,GHz,
4\arcsec~resolution observations of Nagar \etal (2000) show a slightly
resolved  source  with a  total  flux  density  of 7.4~mJy.   Moderate
resolution  observations  at  various  frequencies reveal  a  compact,
flat-spectrum  core (Hummel  \etal  1987; Condon  \etal 1990,  Condon,
Frayer,  \& Broderick  1991a), although  the 0\arcsecpoint2~resolution
observations of Nagar  \etal (2000) give an upper  limit of 2.8~mJy to
any small-scale radio emission at 15\,GHz.

{\it NGC\,4429:} This source was  not detected in the NVSS.  Wrobel \&
Heeschen  (1991)  establish  an  upper  limit of  0.5~mJy  at  5\,GHz,
5\arcsec~resolution,  while the  Nagar  \etal (2000),  0\arcsecpoint2~
resolution  observations  give  an  upper  limit of  1.1~mJy  for  any
small-scale radio emission at 15\,GHz.

{\it NGC\,4435:} NVSS did not  detect this source, while FIRST shows a
2.2~mJy  detection.  5\,GHz, 5\arcsec~resolution  observations (Wrobel
\& Heeschen 1991) detect a weak 1.2~mJy source.  However, the 15\,GHz,
0\arcsecpoint2~resolution observations  of Nagar \etal  (2000) give an
upper limit of 1.1~mJy to any small-scale radio emission.

{\it NGC\,4459:}  There is  no NVSS detection  of this  source.  FIRST
detected a 1.8~mJy source, which  together with the Wrobel \& Heeschen
(1991) 0.8~mJy  detection at  5\,GHz,  5\arcsec~resolution, shows  the
compact and flattish spectrum nature  of the source.

{\it NGC\,4527:} NVSS detected a 178~mJy source, 151~mJy were detected
in the  Green Bank  survey and  FIRST reports a  74~mJy source  -- the
radio     source    displays     resolution     effects.     1.4\,GHz,
54\arcsec~resolution  observations  (Condon  1987)  detect  a  214~mJy
source.   While  0\arcsecpoint2~resolution,  15\,GHz  observations  of
Nagar \etal (2000)  give an upper limit of 1.1~mJy  to any small-scale
emission, the high-resolution observations of Vila \etal (1990) show a
small  ($<$~2\arcsec)  core  (1.1  and  4.2~mJy  at  5  and  1.4\,GHz,
respectively) and East-West extended emission  (73 and 24~mJy at 5 and
1.4\,GHz, respectively).  The  annulus-like extended emission, roughly
aligned with the galaxy major axis, is suggestive of star-formation in
the disk.

{\it NGC\,4569:} NVSS reports a  73~mJy source, while FIRST detected a
14.3~mJy source.   1.4\,GHz, 48\arcsec~resolution observations (Condon
1987;  Condon  \etal  1990,  1998b)  detect  a  83~mJy  source.   High
resolution  2.4\,GHz   observations  show  an   uncertain  3~mJy  core
detection    (Turner,   Helou,   \&    Terzian   1988),    while   the
0\arcsecpoint2~resolution observations  of Nagar \etal  (2000) give an
upper limit of 1.1~mJy to any radio emission at 15\,GHz.

{\it  NGC\,5055:} NVSS  detected a  265~mJy, the  Green Bank  survey a
124~mJy, and FIRST  a 20.2~mJy source.  1.4\,GHz, 54\arcsec~resolution
observations  (Condon   1987)  detect  an   uncertain  22~mJy  source.
Moderate to high-resolution observations at various frequencies either
fail to  detect or detect  uncertain, weak emission from  this object,
denoting  that it  is mainly  low surface  brightness, arcsecond-scale
emission  (Condon  \&  Broderick   1988;  Condon  \etal  1991a).   The
0\arcsecpoint2~resolution observations  of Nagar \etal  (2000) give an
upper limit of 1.1~mJy to any small-scale radio emission at 15\,GHz.

{\it NGC\,5701:} This source was  not detected by the NVSS.  1.4\,GHz,
54\arcsec~resolution  observations (Condon 1987)  also did  not detect
this source.  Hummel \etal (1987)  establish an upper limit of 0.6~mJy
to   any  emission   at   1.4\,GHz,  1\arcsecpoint3~resolution.    The
0\arcsecpoint2~resolution observations  of Nagar \etal  (2000) give an
upper limit of 1.1~mJy to any small-scale radio emission at 15\,GHz.

{\it NGC\,5746:}  NVSS reports  a 19~mJy and  FIRST a  1.2~mJy source.
1.4\,GHz, 54\arcsec~resolution  observations (Condon 1987)  measure an
uncertain 11~mJy  (due to confusion from a  strong, unrelated Northern
source).  The high-resolution observations at 1.4\,GHz of Hummel \etal
(1987) establish an upper limit of 0.8~mJy to the radio core emission.

{\it NGC\,5866:} NVSS detected a 23~mJy source, while FIRST detected a
16.9~mJy  source.  15\arcsec~and  5\arcsec~resolution  observations at
1.4\,GHz   (Condon  \etal   1990)  detect   21  and   12~mJy  sources,
respectively.  Multi-frequency observations of this source suggests it
harbours a  compact, flat-spectrum  radio core (Hummel  1980; Fabbiano
\etal 1987; Wrobel \& Heeschen 1991).  The Nagar \etal (2000) 15\,GHz,
0\arcsecpoint2~resolution   observations  detect  a   7.5~mJy  compact
source, while VLBA  5\,GHz observations of Falcke \etal  (2000) show a
pointlike 8.4\,mJy source.

\subsection {Radio Source Parameters}

We have  used AIPS  task IMFIT in  order to fit  single bi-dimensional
Gaussians to  the $\sim$arcsec-scale brightness peaks  in our sources.
Results  of  this  procedure  are  listed in  Table~4.   We  have,  in
addition,  tabulated the  Gaussian  fits to  the  two strongest,  most
Northeastern  components (NGC\,660C1  and NGC\,660C2)  associated with
NGC\,660.  Given the good phase  stability, we estimate that the radio
peak positions are accurate to within \ltaprx1\arcsec.  The integrated
flux densities  of very  extended sources --  shown in  parentheses --
have    been     estimated    from    the     low-resolution    images
(5\arcsec--7\arcsec~resolution) in combination  with the list of CLEAN
components.

\medskip 
\centerline{\bf -- TABLE~4: Radio parameters of detected sources --}

\section {Discussion and Statistical Results}

Table~5  summarizes  the  intrinsic  properties of  all  65  composite
LINER/H~{\sc ii}  galaxies in  the Palomar sample.   We have  added to
Table~5 12 H~{\sc  ii} galaxies of the same  sample (see Paper~I), for
use as a control group for the statistical properties of the composite
galaxies.    The  H$\alpha$   and  [O~{\sc   i}]   $\lambda$6300  line
luminosities  have   been  measured  in   a  2\arcsec$\times$4\arcsec\
aperture (Ho \etal 1997a), comparable to the resolution of the present
radio data (2.5--5\arcsec). The H$\alpha$ and [O~{\sc i}] entries have
been  corrected for  Galactic and  internal extinction  (see  Ho \etal
1997a) and only photometric values have been considered.

\medskip 
\centerline{\bf -- TABLE~5: summary --}

Since  our aim  is  to study  radio  core emission  in a  homogeneous,
systematic way,  we have used only  high-resolution radio observations
(\ltaprx5\arcsec).   This is generally  accomplished at  8.4\,GHz (our
observations) and  5\,GHz (literature).  We also  attempt to determine
radio spectral  indices for the  compact core components,  using radio
data  at  comparable   resolution.   When  multiple  observations  are
available, radio  data are chosen  by order of preference  as follows:
Paper~I or this  paper, Nagar \etal (2000), Wrobel  \& Heeschen (1991)
and FIRST.   Moreover, in many sources `extended'  emission is present
which is most likely associated with the host galaxy and therefore may
contaminate radio  core flux  densities and morphologies.   In summary
however, combination  of the new observations and  the literature data
permits a census of the occurrence of the $\sim$100 parsec-scale radio
cores  and associated  small-scale structure  in the  composite source
class   of  the  Palomar   bright  galaxy   sample.   We   proceed  by
investigating interesting trends.

In the absence of adequate  empirical constraints, we have assumed the
simplest possible  case, that  is that most  of the radio  emission is
optically thin synchrotron radiation and adopt a spectral index of 0.7
in order  to extrapolate the  radio power entries to  5\,GHz (Table~5;
column~5).  Integrated  flux densities  for NGC\,660, 1055,  3627, and
7331, appearing  in parenthesis in Table~5 (column~5),  have also been
extrapolated to 5\,GHz assuming a spectral index of 0.7.  We note that
an error of $\pm$0.2 in the adopted radio spectral index would lead to
an error of $\pm$0.1 in the logarithm of the extrapolated 5\,GHz radio
power.  Undetected sources  in our observations have all  been given a
conservative  upper   limit  of  0.5~mJy,   corresponding  roughly  to
5$\sigma$.

We have adopted the  radio morphological classification of Ulvestad \&
Wilson  (1984),  applying  this  classification  to  the  $\sim$arcsec
resolution images: the classification refers to radio structure on the
$\sim$100  parsec  scale.  A  source  is  considered  resolved if  the
deconvolved source size is larger  than half the beamwidth in at least
one component.   $U$ and $S$  refer respectively to  single unresolved
and slightly  resolved cores, $D$  is for diffuse radio  emission, $L$
for linear or multicomponent  structure, and $A$ pertains to ambiguous
sources.

Of the 65 composite sources in  the Palomar sample there are 37 (57\%)
arcsec resolution  radio detections,  implying mJy level  structure on
the  $\sim$100  parsec  scale.  Of  these  37,  29  are from  our  VLA
observations  while 8  are  from published  observations.   Of the  37
detections, there are 10 unresolved cores, 14 slightly resolved cores,
3 cores with  associated radio halos, 5 diffuse  sources and 5 sources
with ambiguous radio morphology.  Nine of the twelve H~{\sc ii} nuclei
were detected, seven  of which show diffuse source  morphology and two
of which are ambiguous -- hence no compact cores in this class.

As  in  Paper~I  we  have  divided  the  composite  sources  into  two
categories, based  on their radio morphology and  strength.  The first
category  are  non-AGN composite  sources  in  which  the weak  and/or
extended, low surface brightness radio emission is most likely stellar
in nature.  We  have also included in this  category slightly resolved
core-like    sources   with    peak   flux    densities   \ltaprx1~mJy
($\sim$7$\sigma$).  The  second category  of objects are  AGN-like, in
that they display clear compact  radio cores with flux densities above
1~mJy.  These AGN candidates include NGC\,3627, 4552, 5354, 5838, 5846
(Paper~I) and NGC\,410, 524, 1055,  1161, 3245, 4216, 4419, 4527, 5866
(this paper).  We have added to this category the sources NGC\,660 and
NGC\,7331,  for which  much higher  sensitivity and  higher resolution
observations  are available.   We stress  that both  sources  would be
excluded  from  this  classification  on  the  basis  of  arcsec-scale
observations (see Fig.~2d and 3c).  This illustrates the fact that the
actual fraction of  AGN may be higher than we  derive using the arcsec
resolution data -- an issue we will come back to in Sect.~5.4.

We  have  constructed  binned  distributions  for  the  nuclear  radio
morphology,   radio   spectral   index,   5\,GHz   radio   power   and
extinction-corrected H$\alpha$  luminosity for the  complete composite
source sample -- AGN candidates and  non-AGN sources -- as well as the
12 H~{\sc ii} galaxies (Fig.~4).

\medskip 
\centerline{\bf -- Histograms}

\subsection {Radio Morphology}

The presence of a `prominent' ($>$1~mJy), compact radio core qualifies
a  composite  source  as  an  AGN candidate.   Unresolved  cores  were
detected in  63\% of the  AGN candidates.  In  31\% of the  cases, the
core is  slightly resolved and in  one source, NGC\,4527,  there is in
addition a circumnuclear halo  of diffuse radio emission.  However, no
jet-like emission is detected in  any of the sources and most probably
any extended emission is diffuse galactic emission.

Non-AGN composite sources show two types of morphology.  About 43\% of
the non-AGN  composite sources are  quoted as single  resolved cores.
Some  are classified  as such  based  simply on  visual inspection  of
5\arcsec~resolution FIRST images (NGC\,3898; NGC\,5746); however, they
are all  weak core-like sources (F$_{peak}\ltaprx$1~mJy)  which may be
resolved out  by high-resolution observations (e.g.  Nagar \etal 2000;
Falcke \etal 2000; Nagar \etal  2002; Filho \etal 2002) or may perhaps
fall into  the AGN candidate  category if more  sensitive observations
were  available  (see  discussion  on NGC\,660  and  NGC\,7331).   The
remaining non-AGN composite sources are similar in radio morphology to
the H~{\sc  ii} nuclei: both are  dominated by the  presence of large,
diffuse and/or ambiguous radio emission.

\subsection {Radio Spectral Index}

Because  spectral  index values  are  significantly  affected by  such
diverse  effects as  resolution mismatch,  variability  and underlying
diffuse emission,  we have not  attempted to establish upper  or lower
limits.   The  spectral indices  should  be  considered with  caution,
especially in the case of  H~{\sc ii} nuclei, where the radio emission
is extended.

The  median spectral index  for the  AGN candidate,  non-AGN composite
source  and  H~{\sc  ii}  nuclei  distribution is  0.45,  0.60,  0.40,
respectively.   Grouping all  composite sources  together,  the median
spectral index is 0.5.

With  the exception of  two sources,  all of  the AGN  candidates show
radio   cores   with   relatively   flat/inverted   spectral   indices
($\alpha\leq$0.5).  In  the case of NGC\,3627  and NGC\,4527, extended
disk radio  emission has most  probably contaminated the  radio `core'
flux densities  used in the  spectral index values.  For  NGC\,660, we
argue that  the similarity with  the radio structure in  NGC\,1055 and
NGC\,7331 as well as the coincidence of the detected radio `core' with
the optical  center of the  galaxy and its  radio power, make  it very
unlikely that the `core' is stellar in origin.

Flat  radio spectra can  be produced  by thermal  gas --  by free-free
absorption  of  steep  synchrotron  emission  and  by  optically  thin
free-free emission  -- or by  a synchrotron self-absorbed  radio core.
Nuclear starbursts can have maximum brightness temperatures of T$_{\rm
B}\ltaprx$10$^{4-5}$~K (Condon \etal  1991b).  Although the brightness
temperature     figures     for     arcsec-scale     data     (T$_{\rm
B}\ltaprx$10$^{2-3}$~K) do  not permit  us to disentangle  the thermal
versus nonthermal synchrotron emission,  VLBA observations of 4 out of
6 AGN  candidates (Falcke  \etal 2000; Nagar  \etal 2002;  Filho \etal
2002) show  that these  sources do  in fact  harbour  nonthermal, high
brightness temperature  (T$_{\rm B} \gtaprx  $10$^6$~K), compact radio
cores,  consistent with  the presence  of a  synchrotron self-absorbed
source.

Most non-AGN  composite sources show a tendency  for slightly steeper
spectral  indices ($\alpha\geq$0.5),  typical  of nonthermal  sources,
like  supernova remnants  (e.g.  Tarchi  \etal 2000).   Another subset
shows  relatively flat  spectral indices  ($\alpha\leq$0.5) associated
with either diffuse radio  morphology or weak core-like sources.  Most
likely these are objects whose  radio emission is dominated by thermal
processes, similar to that occuring  in H~{\sc ii} nuclei.  All H~{\sc
ii}  nuclei have  flattish ($\alpha\sim$0.4)  spectra as  expected for
thermal emission  except for NGC\,4800, whose steep  spectral index is
probably  a  result  of   resolution  effects  or  contamination  from
nonthermal radio emission.

\subsection {Radio Power}

Some of  the radio  luminosities are censored;  only upper  limits are
available.   In order  to  treat such  a  censored data  set, we  have
applied the Kaplan-Meier  product-limit estimator (Feigelson \& Nelson
1985).   The results show  that the  mean radio  power for  the arcsec
scale core emission in  AGN candidates, non-AGN composite sources and
H~{\sc  ii}  nuclei  is respectively,  (1.22$\pm$0.58)$\times10^{20}$,
(1.19$\pm$0.43)$\times10^{20}$,  and (5.28$\pm$1.37)$\times10^{19}$ in
units of  W Hz$^{-1}$.  Grouping all composite  sources together, the
mean is (1.20$\pm$0.35)$\times10^{20}$ W Hz$^{-1}$.

We see no statistical difference between the three classes of sources:
their  radio  power distributions  are  similar.   In particular,  the
composite  source nuclei  populate the  lower end  of  the integrated
luminosity   range  of  `normal'   nearby  galaxies   (10$^{18-23}$  W
Hz$^{-1}$; see, e.g., Condon 1987,  1992) and the least luminous radio
cores in Seyfert galaxies (10$^{19-22}$  W Hz$^{-1}$; see, e.g., Ho \&
Ulvestad  2001;  Ulvestad \&  Ho  2001).   Moreover, these  composite
sources  are extremely  faint when  compared to  traditionally studied
AGN: a typical  FR\,I radio galaxy core has  a 5\,GHz radio luminosity
in the range 10$^{21-25}$ W Hz$^{-1}$ (Zirbel \& Baum 1995).

\subsection {Radio Continuum and Line Emission}

Active   galaxies   display   correlated   radio   and   emission-line
luminosities; these are well studied in Seyferts (Ho \& Ulvestad 2001;
Ulvestad \& Ho 2001), radio galaxies (Zirbel \& Baum 1995) and also in
LINERs (Ho 1999; Nagar \etal 2000, 2002).

The  median  extinction-corrected  H$\alpha$  luminosity for  the  AGN
candidates,  non-AGN  composite  sources  and H~{\sc  ii}  nuclei  is,
respectively, 10$^{39.01}$, 10$^{38.82}$, and 10$^{39.59}$ in units of
erg  s$^{-1}$.  Taking  all composite  sources together,  we  obtain a
median of 10$^{38.87}$ erg s$^{-1}$.

Although  the peak  in the  H~{\sc  ii} nuclei  distribution hints  at
slightly larger  H$\alpha$ luminosities, on average,  we conclude that
there are no significant  differences in the H$\alpha$ luminosities of
the three classes of sources.

In order to ascertain the nature of the radio emission and investigate
any  systematic trends between  the different  classes, we  will start
from the simplistic  premise that all the H$\alpha$  luminosity is due
to  stellar  processes.  Case  B  recombination  theory  allows us  to
calculate the  number of  ionizing photons needed  to produce  a given
H$\alpha$ luminosity,  which in turn  allows us to calculate  how much
free-free radio  emission should be  produced at a  certain frequency.
For  a  purely  thermal  source  of  T$_{e^-}\sim$10$^4$~K  and  given
H$\alpha$ luminosity,  Case B  recombination predicts a  certain radio
power at  5\,GHz, given approximately by (Ulvestad,  Wilson, \& Sramek
1981):

\begin{center}
$ {\rm P_{5\,GHz}^{predicted}} = 10^{-20} {\rm L(H\alpha)}
\hspace{2cm} {\rm [W \, Hz^{-1}]} $
\end{center}

\noindent where L(H$\alpha$) is expressed in erg s$^{-1}$.

We  have  plotted  in  Fig.~5  the  nuclear  radio  power  versus  the
extinction-corrected  H$\alpha$  luminosity  for the  AGN  candidates,
non-AGN  composite sources  and H~{\sc  ii} nuclei.   The  solid line
defines  the  predicted  thermal  radio  power if  all  the  H$\alpha$
luminosity  were  due  to   star  formation.   Plotted  also  are  the
best-fitting linear  correlations found  for FR I  (dashed) and  FR II
(dotted) radio galaxy cores (Zirbel \& Baum 1995).

\medskip 
\centerline{\bf -- Plot P$_{\rm 5\,GHz}$ vs. L$_{\rm H\alpha}$}

H~{\sc  ii} galaxies  seem to  follow the  thermal relation  slope (to
within  \ltaprx0.5  order of  magnitude)  with  slightly higher  radio
luminosities  than   predicted.   However,  at   any  given  H$\alpha$
luminosity there is a range of almost two orders of magnitude in radio
power in  all composite  sources (see also  Nagar \etal 2002).   If we
compare  our  plot with  similar  diagrams  constructed for  Seyferts,
LINERs and  radio galaxies, we see that,  despite substantial scatter,
our composite sources seem to follow the more powerful AGN trends down
to lower H$\alpha$ luminosities and  radio powers (Ho 1999; Ho \& Peng
2001; Nagar \etal 2002).

We have investigated the possible correlation between P$_{\rm 5\,GHz}$
and L$_{\rm  H\alpha}$ using the  generalized Kendall $\tau$  test for
censored  data  (Isobe, Feigelson,  \&  Nelson  1986), where  $\tau=1$
implies direct correlation, $\tau=-1$ inverse correlation and $\tau=0$
no correlation.   Censoring was considered in  the dependent variable:
radio  power.    Given  the  small  number  statistics   for  the  AGN
candidates,  we have  analysed  AGN candidates  and non-AGN  composite
sources   together.   The   computed  generalized   Kendall's   $\tau$
correlation  coefficient  for composite  sources  and  H~{\sc ii}  is,
respectively, 0.18 and 0.51, and  the $Z$-value is 1.25 and 1.39.  The
probability   that  P$_{\rm   5\,GHz}$  and   L$_{\rm   H\alpha}$  are
uncorrelated is 21\% and 17\%, respectively, for composite sources and
H~{\sc ii} nuclei.

Most  of  the  composite  sources  fall  above  the  predicted  5\,GHz
threshold for  thermal free-free emission, indicating  that if thermal
emission is present,  it does not dominate the  radio continuum. There
must be an additional nonthermal component, not necessarily an AGN but
possibly  supernova  remnants, contributing  to  the radio  continuum.
Interestingly, three  of the  AGN candidates (NGC\,660,  NGC\,4527 and
NGC\,7331) having  measured radio core resolutions about  a factor two
below  the  angular  scale   region  used  to  measure  the  H$\alpha$
luminosity,  fall rather low  with respect  to the  main body  of data
points.

Combining both samples in Paper~I  and the present paper, we obtain an
overall  detection  rate of  radio  cores  in  25\% of  the  composite
LINER/H~{\sc ii} galaxies, compared to the 17\% detected in a distance
limited sample of composite sources  (Nagar \etal 2000, 2002). We find
that radio cores  occur not only in the ellipticals  of our sample but
also in the bulge-dominated  spirals, confirming the trend reported in
Paper~I (see also  Nagar \etal 2002). This detection  rate is a factor
two lower than in genuine LINERs (Nagar \etal 2000, 2002) and a factor
three  lower  than  the detection  rate  of  radio  cores in  the  low
luminosity  Seyfert galaxies  of the  Palomar sample  (Ho  \& Ulvestad
2001), even  though the  sensitivity, resolution and  radio luminosity
range are  similar.  Now, lower  radio core luminosities  in composite
sources are not unexpected since the core luminosity in LINERs and low
luminosity Seyferts correlates with the [O~{\sc i}] $\lambda$6300 line
luminosity, and this neutral  oxygen luminosity is significantly lower
in composite  sources (Nagar  \etal 2000, 2002).   Therefore, although
many composite  sources may  harbour an AGN  core, only  the strongest
radio cores will be detected.

We  have  plotted  in  Fig.~6  the  nuclear  radio  power  versus  the
extinction-corrected  [O~{\sc i}] line  luminosity (using  the [O~{\sc
i}] to H$\alpha$ line ratio in column~(4), Table~6).

\medskip 
\centerline{\bf -- Plot P$_{\rm 5\,GHz}$ vs. L$_{\rm _{[O~{\sc i}]}}$}

As with the H$\alpha$  luminosity, the correlation between radio power
and [O~{\sc  i}] line  emission is absent,  with radio  power spanning
almost two orders  of magnitude at a given  [O~{\sc i}] luminosity for
all three different source types.

The  significance  of  the   correlation  can  be  quantified  by  the
generalized Kendall $\tau$ test, where again non-AGN composite sources
and  AGN candidates have  been grouped  together.  We  have considered
upper  limits in  both  the dependent  (radio  power) and  independent
variable   ([O~{\sc  i}]  line   luminosity).   The   computed  $\tau$
correlation coefficient  for composite sources and H~{\sc  ii} is 0.07
and 0.18, and the $Z$-value 0.54 and 0.52, respectively.  This implies
that  the   probability  of  P$_{\rm  5\,GHz}$   and  L$_{\rm  [O~{\sc
i}]\lambda6300}$  being uncorrelated  is 59\%  and 61\%  for composite
sources and H~{\sc ii} nuclei, respectively.

In  conclusion,  compact,   flat/inverted  spectrum  radio  cores  are
detected in almost  a quarter of all the  composite objects, existing
not only  in ellipticals but  also in bulge-dominated  spirals (Sa-Sb;
see also Nagar \etal 2002).   Curiously, only one elliptical galaxy in
our complete  composite source sample  -- NGC\,4125 -- shows  no radio
evidence, on any scale, of an active nucleus (see Section~4.2).  While
a spheroid seems  to be a necessary condition for  AGN activity, it is
certainly not a sufficient one.

On the other hand, composite  objects, even the AGN candidates, do not
seem  to  follow the  radio  versus  [O~{\sc  i}] and  H$\alpha$  line
luminosity correlations  seen in classical AGN and  low luminosity AGN
like  LINERs (Ho \&  Ulvestad 2001;  Nagar \etal  2002).  The  lack of
correlation may be related to  the resolution -- at arcsec resolution,
the radio emission may still be dominated by thermal processes related
to  circum-nuclear  star   formation  (see  NGC\,660  and  NGC\,7331).
However, except in their radio properties, we see no systematic trends
that distinguish  non-AGN composite  sources from the  AGN candidates.
In fact, scaling arguments suggest that even composite sources without
detected radio cores  could be lower power versions  of our radio core
detections.  This  suggests that there  is no obvious reason  for {\it
all} composite sources {\it not} to  harbour an AGN core.  We may only
be detecting the high end of the composite source radio power, whereby
weaker radio  cores are  drowned out by  more dominating  star forming
regions (see discussion  on NGC\,660 and NGC\,7331).  The  clue may be
in  resolution and  sensitivity  --  detecting a  very  weak AGN  core
requires sensitive  high dynamic range imaging. A  complete AGN census
requires a sensitive VLBI survey.

\section{Conclusions}

We have  embarked on  a radio  imaging study of  a complete  sample of
composite sources from the  Palomar magnitude-limited sample of nearby
emission-line  nuclei.   Our  results  show  that  a  quarter  of  all
composite  sources host  faint AGN,  as  revealed by  the presence  of
compact, flat/inverted spectrum radio cores.  Furthermore, these cores
occur  not only in  elliptical galaxies,  but also  in bulge-dominated
spirals.   The  rather low  radio  core  luminosities  suggest we  are
probing the very faint end  of the AGN luminosity function.  We cannot
as yet exclude the possibility of a higher AGN fraction.

\acknowledgments

M.~E.~F.   is  supported  by  grant PRAXIS  XXI/BD/15830/98  from  the
Funda\c c\~ao para a Ci\^encia e Tecnologia, Minist\'erio da Ci\^encia
e Tecnologia, Portugal.  L.~C.~H.  is  partly funded by NASA grant NAG
5-3556, and  by NASA  grants GO-06837.01-95A and  AR-07527.02-96A from
the Space  Telescope Science Institute (operated by  AURA, Inc., under
NASA contract NAS5-26555).  We have made extensive use of the NVSS and
FIRST online database,  as well as online VLA  calibrator manuals.  We
want to thank  Jim Ulvestad for providing us  with calibrated VLA data
and Leticia Martin-Hernandez for some useful assistance.

We  acknowledge the VLA,  which is  a facility  of the  National Radio
Astronomy  Observatory (NRAO)  which  is a  facility  of the  National
Science Foundation operated  under cooperative agreement by Associated
Universities, Inc.  We  have made extensive use of  the APM (Automatic
Plate  Measuring)  Facility, run  by  the  Institute  of Astronomy  in
Cambridge,  the STScI DSS  (Digitized Sky  Survey), produced  under US
government  grant  NAGW  --  2166, and  NED  (NASA/IPAC  Extragalactic
Database),  which  is  operated  by  the  Jet  Propulsion  Laboratory,
California Institute of Technology, under contract with NASA.

\clearpage

\footnotesize
\begin{deluxetable}{lcccrcl}
\tablenum{1}
\tablecolumns{7} 
\tablewidth{0pc}
\tablecaption{Composite source sample. 
Col. 1 and 2: Source name. 
Col. 3 and 4: Optical position from NED (NASA/IPAC Extragalactic Database).
Col. 5:  Adopted distance from Tully 1988, with $H_{\rm 0}$=75 km s$^{-1}$ Mpc$^{-1}$.
Col. 6:  Hubble  type from NED.
Col. 7: References to relevant earlier  radio data.}
\tablehead{ 
\colhead{} & \colhead{Other} & \colhead{R.A.} & \colhead {Decl.} & \colhead{$D$} & \colhead {} 
& \colhead{} \\
\colhead{Galaxy} & \colhead{Name} & \colhead{(J2000)} & \colhead{(J2000)} & \colhead{(Mpc)} & 
\colhead {Hubble Type} & \colhead{Reference} \\
\colhead{(1)} & \colhead {(2)} & \colhead{(3)} & \colhead{(4)} & \colhead{(5)} & \colhead{(6)} & \colhead{(7)}}
\startdata 

IC\,356\tablenotemark{a}& & 04 07 46.8 & $+$69 48 45 & 18.1 & SA(s)ab pec & 1,2 \\

IC\,1727\tablenotemark{a} & & 01 47 31.3 & $+$27 19 39 & 8.2 & SB(s)m & 2,3 \\

NGC\,410\tablenotemark{a} & & 01 10 58.9 & $+$33 09 08 & 70.6 & E+: & 3,4  \\

NGC\,488\tablenotemark{a} & & 01 21 46.8 & $+$05 15 25 & 29.3 & SA(r)b & 2,3,5,6,7 \\

NGC\,521\tablenotemark{a} & & 01 24 33.8 & $+$01 43 52 & 67.0 & SB(r)bc& 3,8 \\

NGC\,524\tablenotemark{a} & & 01 24 47.8 & $+$09 32 19 & 32.1 & SA(rs)0 + & 4,6,7,9,10,11  \\

NGC\,660 & & 01 43 01.7 & $+$13 38 34 & 11.8   & SB(s)a pec & 1,2,3,4,5,12,13 \\
       & &            &           &          &              & 14,15,16,17,18\\
       & &            &           &          &              & 19,20,21,22,23,24\\

NGC\,1055\tablenotemark{a} & & 02 41 45.2 & $+$00 26 30 & 12.6 & SBb: spin & 1,2,3,4,5,6,7,12 \\
        & &            &           &          &                            & 15,16,19,25,26,27\\

NGC\,1161\tablenotemark{a}  &  & 03 01 14.1 & $+$44 53 49 & 25.9 & SA0 & 4  \\

NGC\,2859 & & 09 24 18.5 & $+$34 30 49 & 25.4 & (R)SB(r)0 + & 3,7,9,10,11,28 \\

NGC\,3245 & & 10 27 18.4 & $+$28 30 27 & 22.2 & SA(r)0? & 4,7,9,11,25 \\

NGC\,3489 & & 11 00 18.2 & $+$13 54 06 & 6.4 & SAB(rs)0 + & 3,6,7,9,10,11 \\

NGC\,3692 & & 11 28 24.2 & $+$09 24 23 & 29.8 & SAb & 3,4,29  \\

NGC\,3705 & & 11 30 06.7 & $+$09 16 36 & 17.0 & SAB(r)ab & 1,2,3,6,7,29  \\

NGC\,3898 & & 11 49 15.1 & $+$56 05 01 & 21.9 & SA(s)ab & 2,4,7,25,28,30,31 \\
        & &            &           &        &         &  32,33  \\

NGC\,3917 & & 11 50 45.4 & $+$51 49 28 & 17.0 & SAcd: & 7,29,34  \\

NGC\,3953 & & 11 53 49.0 & $+$52 19 37 & 17.0 & SB(r)bc & 1,2,7,8,15,26  \\
          & &            &           &      &           & 28,29,32,33,35    \\

NGC\,3992 & M\,109 & 11 57 36.1 & $+$53 22 29 & 17.0 & SB(rs)bc &  1,2,7,8,28,29,32\\ 

NGC\,4125 & & 12 08 05.7 & $+$65 10 24 & 24.2 & E6 pec & 7,9,32,36,37 \\

NGC\,4145 & & 12 10 01.6 & $+$39 52 59 & 20.7 & SAB(rs)d & 2,7,29 \\

NGC\,4150 & & 12 10 33.3 & $+$30 24 12 & 9.7 & SA(r)0 ? & 3,7,9,11,36 \\

NGC\,4192 & M\,98 & 12 13 48.3 & $+$14 54 01 & 16.8 & SAB(s)ab & 1,2,3,4,5,6,7,12 \\
        &       &            &           &           &       & 13,15,19,25,29,38 \\

NGC\,4216 & & 12 15 54.2 & $+$13 08 59 & 16.8 & SAB(s)b: & 1,2,3,4,5,6,7 \\
          & &            &           &      &          & 15,29 \\

NGC\,4220 & & 12 16 11.8 & $+$47 52 58 & 17.0 &  SA(r)0 + & 4,9,11,26,29  \\

NGC\,4281 & & 12 20 21.5 & $+$05 23 11 & 35.1 & SA0+: spin & 3,6,9,11  \\

NGC\,4324 & & 12 23 06.1 & $+$05 15 00 & 35.1 & SA(r)0 + & 3,9  \\

NGC\,4350 & & 12 23 57.5 & $+$16 41 35 & 16.8 & SAO spin & 3,7,9,11 \\

NGC\,4419 &  & 12 26 56.4 & $+$15 02 48 & 16.8 & SB(s)a spin & 3,4,5,12,13,15,24,29  \\

NGC\,4429 & & 12 27 26.4 & $+$11 06 29 & 16.8 & SA(r)0 + & 3,6,7,9,11,29 \\

NGC\,4435 & & 12 27 40.5 & $+$13 04 44 & 16.8 & SB(s)0 & 3,7,9,11,25,29,39,40  \\

\enddata
\end{deluxetable}

\clearpage

\footnotesize
\begin{deluxetable}{llccrcl}
\tablenum{1}
\tablecolumns{7} 
\tablewidth{0pc} 
\tablecaption{Sample sources (cont.)
}
\tablehead{ 
\colhead{} & \colhead{Other} & \colhead{R.A.} & \colhead {Decl.} & \colhead{$D$} & \colhead {} 
& \colhead{} \\
\colhead{Galaxy} & \colhead{Name} & \colhead{(J2000)} & \colhead{(J2000)} & \colhead{(Mpc)} & 
\colhead {Hubble Type} & \colhead{Reference} \\
\colhead{(1)} & \colhead {(2)} & \colhead{(3)} & \colhead{(4)} & \colhead{(5)} & \colhead{(6)} & \colhead{(7)}}
\startdata 

NGC\,4459 & & 12 28 59.9 & $+$13 58 45 & 16.8 & SA(r)0 + & 3,9,10,11,25 \\

NGC\,4527 & & 12 34 08.5 & $+$02 39 11 & 13.5 & SAB(s)bc & 1,2,3,4,5,6,7,8,12,15 \\
        & &            &           &          &        & 16,25,29,41 \\

NGC\,4569 & M\,90 & 12 36 49.8 & $+$13 09 46 & 16.8 & SAB(rs)ab & 1,2,3,4,5,6,7,9,12,13 \\
        &       &            &           &           &          & 15,17,25,29,36,38,39 \\
        &       &            &           &           &          & 42,43\\

NGC\,5055 & M\,63 & 13 15 49.2 & $+$42 01 49 & 7.2 & SA(rs)bc & 1,2,4,7,8,12,13,15,16 \\
        &       &            &           &          &         & 19,24,25,26,28,29,32 \\
        &       &            &           &          &         & 42,44,45,46 \\

NGC\,5701 & & 14 39 11.2 & $+$05 21 56 & 26.1 & (R)SB(rs)0/a & 2,3,5,6,29  \\

NGC\,5746 & & 14 44 56.4 & $+$01 57 16 & 29.4 & SAB(rs)b? spin & 2,4,5,6,25,29  \\

NGC\,5866 & M\,102 & 15 06 29.4 & $+$55 45 49 & 15.3 & SA0 + spin & 4,7,9,10,11,15,25 \\
        &        &            &           &          &          & 29,32,47,48,49  \\

NGC\,7177\tablenotemark{a} & & 22 00 41.6 & $+$17 44 17 & 18.2 & SAB(r)b & 1,2,3,4,5 \\

NGC\,7331\tablenotemark{a} & & 22 37 04.1 & $+$34 24 56 & 14.3 & SA(s)b & 1,2,3,7,8,13,16,19,24 \\
        & &            &           &          &                         & 26,28,44,48,50,51,52 \\

NGC\,7742\tablenotemark{a} & & 23 44 15.9 & $+$10 46 04 & 22.2 & SA(r)b & 1,4,29  \\

\enddata

\tablecomments {Units  of right ascension  are hours, minutes, and  seconds, and
units of declination are degress, arcminutes, and arcseconds.}

\tablenotetext{a}{Galaxies with new VLA observations.}

\tablerefs {(1)  Niklas \etal  1995; (2) Condon  1987; (3)  Dressel \&
Condon 1978; (4) Condon \etal 1998a (NVSS); (5) Hummel \etal 1987; (6)
Harnett 1982;  (7) Hummel 1980; (8)  Hummel \etal 1985;  (9) Wrobel \&
Heeschen 1991; (10) Fabbiano \etal  1987; (11) Hummel \& Kotanyi 1982;
(12) Condon  \etal 1995; (13)  Condon \etal  1991a; (14)  Carral \etal
1990;  (15) Condon  \etal 1990;  (16) Condon  \& Broderick  1988; (17)
Wunderlich \etal 1987; (18) Hummel  \etal 1984; (19) Israel \& van der
Hulst 1983; (20) Condon \etal 1982; (21) Jones \etal 1981; (22) Condon
1980; (23) Condon \& Dressel 1978; (24) Sramek 1975; (25) Becker \etal
1995 (FIRST);  (26) Israel  \& Mahoney 1990;  (27) Downes  \etal 1986;
(28) Gioia \&  Gregorinni 1980; (29)  Nagar \etal 2000; (30)  Gioia \&
Fabbiano 1987; (31) Brosch \&  Krumm 1984; (32) Heckman 1980; (33) van
der  Kruit 1971;  (34) Condon  \etal 1987;  (35)  Garcia-Barreto \etal
1993; (36)  Condon \etal  1998b; (37) Hummel  \etal 1983;  (38) Turner
\etal 1988; (39)  Sadler \etal 1995; (40) Kotanyi  \& Ekers 1983; (41)
Vila \etal 1990; (42) Neff \& Hutchings 1992; (43) Urbanik \etal 1986;
(44) Hummel \&  Bosma 1982; (45) Klein  \& Emerson 1981;  (46) van der
Kruit  1973a; (47)  Feretti \&  Giovannini  1980; (48)  van der  Kruit
1973b; (49) Falcke \etal 2000; (50) Dumke \etal 1995; (51) Cowan \etal
1994; (52) Klein \etal 1984.}
                                      
\end{deluxetable}

\clearpage

\footnotesize
\begin{deluxetable}{lrcrcc}
\tablenum{2}
\tablecolumns{6} 
\tablewidth{0pc} 
\tablecaption{Map parameters of the detected sources. Re-reduced source, NGC\,660, is
also included. 
Col. 1: Source name.
Col. 2: Applied taper.
Col. 3: Restoring  beam. 
Col. 4: Position angle  of the  beam.
Col. 5: Rms noise level of the image.
Col. 6: Figure number. 
}
\tablehead{ 
\colhead{} & \colhead{Taper} & \colhead{Beam Size} & \colhead {P.A.} & \colhead{rms} & \colhead {} \\
\colhead{Galaxy} & \colhead{(K$\lambda$)} & \colhead{(arcsec$^2$)} & \colhead{(deg)} & 
\colhead{(mJy beam$^{-1}$)} & \colhead{Figure Number} \\
\colhead{(1)} & \colhead {(2)} & \colhead{(3)} & \colhead{(4)} & \colhead{(5)} & \colhead{(6)}}
\startdata 

IC\,356 & 0 & 3.50 $\times$ 1.98 & 60.00 & 0.081 & 1a \\

IC\,1727 & 0 & 2.47 $\times$ 2.31 & 55.89 & 0.058 & \nodata \\

NGC\,410 & 0 & 2.73 $\times$ 2.30 & 81.97 & 0.043 & 1b \\

NGC\,488 & 0 & 2.93 $\times$ 2.52 & $-$39.51 & 0.049 & 1c\\

NGC\,521 & 0 & 3.04 $\times$ 2.50 & $-$34.01 & 0.046 & 1d \\

NGC\,524 & 0 & 2.66 $\times$ 2.47 & $-$35.18 & 0.043 & 2a \\ 

NGC\,660 & 0 & 0.21 $\times$ 0.21 & $-$39.90 & 0.012 & 2c \\
          
          & 30 & 2.54 $\times$ 2.19 & 67.11 & 0.170 & 2d \\  

NGC\,1055 & 0 & 3.45 $\times$ 2.57 & $-$41.65 & 0.047 & 2b  \\ 

          & 30 & 8.25 $\times$ 6.82 & 21.34 & 0.065 & \nodata \\

NGC\,1161 & 0 & 2.66 $\times$ 2.10 & 44.97 & 0.070 & 3a \\

NGC\,7177 & 0 & 2.40 $\times$ 2.26 & 16.91 & 0.042 & 3b \\

NGC\,7331 & 0 & 2.37 $\times$ 2.15 & 32.26 & 0.039 & 3c \\

          & 30 & 7.64 $\times$ 6.24 & 37.86 & 0.051 & \nodata \\

NGC\,7742 & 0 & 2.52 $\times$ 2.38 & $-$18.76 & 0.039 & 3d \\

          & 30 & 5.68 $\times$ 5.13 & 30.86 & 0.063 & \nodata \\

\enddata
\end{deluxetable}

\clearpage

\footnotesize
\begin{deluxetable}{lccrrr}
\tablenum{3}
\tablecolumns{6} 
\tablewidth{344.3pt} 
\tablecaption{Field source parameters. 
Col. 1: Field source, designated by sky orientation with respect to the field center for the relevant galaxy.
Col. 2: NVSS source name, when available.
Col. 3 and 4: 8.4\,GHz radio position.
Col. 5: NVSS 1.4\,GHz flux density, 45\arcsec~resolution.  
Col. 6: Nuclear 8.4\,GHz flux density. 
}
\tablehead{ 
\colhead{} & \colhead{} & \colhead{R.A.} & \colhead {Decl.} & \colhead{NVSS} & \colhead {$F_{\rm 8.4}^{\rm int}$\tablenotemark{a}}  \\
\colhead{Field Source} & \colhead{Name} & \colhead{(J2000)} & \colhead{(J2000)} & \colhead{(mJy)} & 
\colhead {(mJy)}  \\
\colhead{(1)} & \colhead {(2)} & \colhead{(3)} & \colhead{(4)} & \colhead{(5)} &  \colhead{(6)}}
\startdata 

IC\,356NE  & \nodata & 04 08 21.39 & $+$69 50 04.6 & \nodata & 6.0  \\

NGC\,521SE & J012437+014112 & 01 24 37.97 & $+$01 41 12.0 & 17.6 & 3.5  \\

NGC\,521NE & J012443+014535 & 01 24 43.95 & $+$01 45 36.6 & 5.7  & 1.5  \\

NGC\,524SW &  \nodata & 01 24 41.71 & $+$09 31 20.9 & \nodata & 0.8   \\

NGC\,1161S &  \nodata  & 03 01 13.53 & $+$44 51 59.4 & \nodata & 1.2  \\

IC\,1727SW &\nodata & 01 47 21.55 & $+$27 18 53.3 &\nodata & 0.6   \\

NGC\,7331SW & \nodata & 22 36 54.15 & $+$34 21 14.3 & \nodata & 19.3   \\

NGC\,7742SW & \nodata & 23 44 10.08 & $+$10 45 25.4 & \nodata & 0.8  \\

\enddata 

\tablecomments{Units  of right ascension  are hours, minutes, and  seconds, and
units of declination are degress, arcminutes, and arcseconds.} 

\tablenotetext{a}{The integrated 8.4\,GHz flux densities have been corrected for primary-beam attenuation.}

\end{deluxetable}

\clearpage

\footnotesize
\begin{deluxetable}{lrcccccr}
\tablenum{4}
\tablecolumns{8} 
\tablewidth{0pc} 
\tablecaption{The 8.4\,GHz parameters of the detected radio cores.
Col. 1: Source name.
Col. 2: Applied taper.
Col. 3: 8.4\,GHz peak flux density.
Col. 4 and 5: 8.4\,GHz radio position.
Col. 6: Nuclear 8.4\,GHz flux density.
Col. 7: Deconvolved source size.
Col. 8: Position angle of the source.
}
\tablehead{ 
\colhead{} & \colhead{Taper} & \colhead{$F_{\rm max}$} & \colhead {R.A.} & \colhead{Decl.} & \colhead {$F_{\rm int}$\tablenotemark{a}} 
& \colhead{Size} & \colhead{P.A.} \\
\colhead{Galaxy} & \colhead{(K$\lambda$)} & \colhead{(mJy beam$^{-1}$)} & \colhead{(J2000)} & \colhead{(J2000)} & 
\colhead{(mJy)} & \colhead {(arcsec$^2$)} & \colhead{(deg)} \\
\colhead{(1)} & \colhead {(2)} & \colhead{(3)} & \colhead{(4)} & \colhead{(5)} & \colhead{(6)} & \colhead{(7)} 
& \colhead{(8)}}
\startdata 

IC\,356 & 0 & 0.82 & 04 07 46.67 & 69 48 45.0 & 1.16 & 2.89 $\times$ 0.36 & 34.16  \\

NGC\,410 & 0 & 1.66 & 01 10 58.91 & 33 09 07.0 & 1.75 & 1.19 $\times$ 1.19 & \nodata \\

NGC\,488 & 0 & 0.31 & 01 21 46.77 & 05 15 24.2 & 0.39 & 2.82 $\times$ 2.82 & \nodata \\

NGC\,521 & 0 & 0.40 & 01 24 33.77 & 01 43 53.2 & 0.53 & 2.16 $\times$ 1.04 & 127.57 \\

NGC\,524 & 0 & 1.95 & 01 24 47.75 & 09 32 20.1 & 2.10 & 1.02 $\times$ 0.17 & 115.37 \\ 

NGC\,660 & 0 & 0.43 & 01 43 02.32 & 13 38 44.9 & 3.18 & 0.60 $\times$ 0.47 & 148.82 \\

                & 30 & \nodata & \nodata & \nodata & (60) & \nodata & \nodata \\

NGC\,660C1      & 0 & 0.99 & 01 43 02.40 & 13 38 46.2 & 5.53 & 0.62 $\times$ 0.31 & 59.00 \\

 NGC\,660C2      & 0 & 2.23 & 01 43 02.44 & 13 38 46.4 & 5.36 & 0.28 $\times$ 0.21 & 41.90 \\

NGC\,1055 & 0 & 1.64 & 02 41 45.19 & 00 26 38.3 & 6.71 & 5.79 $\times$ 4.69 & 116.50 \\ 

           & 30 & \nodata & \nodata & \nodata & (40) & \nodata & \nodata \\

NGC\,1161 & 0 & 3.98 & 03 01 14.14 & 44 53 50.51 & 4.25 & 0.65 $\times$ 0.56 & 22.82  \\

NGC\,7177 & 0 & 0.26 & 22 00 41.19 & 17 44 16.2 & 0.82 & 5.13 $\times$ 2.00 & 35.36 \\

NGC\,7331  & 30 & \nodata & \nodata & \nodata  & (50)   & \nodata & \nodata   \\

NGC\,7742 & 30 & 0.49 & 23 44 15.84 & 10 46 01.1 & (21) & 37 $\times$ 32 & 118.99 \\

\enddata
          
\tablecomments{Units  of  right  ascension  are  hours,  minutes,  and
seconds,  and  units  of  declination  are  degress,  arcminutes,  and
arcseconds.}

\tablenotetext{a}{Total flux densities of very extended sources appear
in parentheses.}

\end{deluxetable}

\clearpage

\footnotesize
\begin{deluxetable}{lcrrrrrcrr}
\tablenum{5}
\tablecolumns{10} 
\tablewidth{0pc} 
\tablecaption{Summary of the complete composite galaxy sample.
Col. 1: Source name.
Col. 2: Radio morphology.
Col. 3: H$\alpha$ luminosity corrected for extinction, from Ho \etal 1997a and updated unpublished values. Non-photometric values are not tabulated.
Col. 4: [O~{\sc i}] to H$\alpha$ line ratio, from Ho \etal 1997a.
Col. 5: Logarithm of the monochromatic nuclear radio power.
Col. 6: Reference to the radio morphology and luminosity.
Col. 7: Nuclear radio spectral index.
Col. 8: Reference for the spectral index.
}
\tablehead{ 
\colhead{} & \colhead{} & \colhead{Log L(H$\alpha$)} & \colhead{} & \colhead{log P$_{\rm 5\,GHz}$\tablenotemark{b}} & \colhead{} & \colhead {} & \colhead{}\\
\colhead{Galaxy} & \colhead{Morph.\tablenotemark{a}} & \colhead{(erg s$^{-1}$)} & \colhead{$\frac{[O~{\sc i}]}{H\alpha}$}  & \colhead{(W Hz$^{-1}$)} & \colhead{Ref.} & \colhead{$\alpha_{\rm radio}$} & \colhead{Ref.}\\
\colhead{(1)} & \colhead {(2)} & \colhead{(3)} & \colhead{(4)} & \colhead{(5)} & \colhead{(6)} & \colhead{(7)} & \colhead{(8)}}
\startdata

IC\,356  & S & \nodata & 0.11 & 19.95 & 1 & \nodata & \nodata  \\

IC\,520 & \nodata & 38.92 & 0.097 & $<$20.42 & 2 & \nodata & \nodata \\

IC\,1727 & \nodata & 38.17 & 0.016 & $<$18.90 & 1 & \nodata & \nodata \\

NGC\,410 & S & 39.43 & $<$0.097 & 20.86 & 1 & \nodata  & \nodata \\

NGC\,488 & S & 39.20 & $<$0.13 & 19.90 & 1 & \nodata & \nodata \\

NGC\,521 & S & 39.16 & 0.086 & 20.75 & 1 & \nodata & \nodata \\

NGC\,524 & U & 38.58 & $<$0.14 & 20.26 & 1 & $-$0.8 & 1,3 \\ 

NGC\,660 & S & 40.41 & 0.047 & 19.57 & 1 &  \nodata & \nodata  \\

         & S+L &     &       & (21.30) & 1 & \nodata & \nodata  \\ 

NGC\,1055 & S   & 37.92 & 0.064 & 19.95  & 1 & 0.0 & 1,4 \\ 

          & S+L &       &       & (21.18) & 1 & 1.1 & 1,10\\

NGC\,1161 & U & 39.01 & $<$0.14 & 20.38 & 1 & \nodata & \nodata \\

NGC\,2541 & \nodata & 37.26 & 0.082 & $<$19.12 & 2 & \nodata & \nodata \\

NGC\,2859 & \nodata & 38.59 & 0.092 & $<$19.59 & 3 & \nodata & \nodata \\

NGC\,2985 & S & 38.36 & 0.12 & 20.51 & 2 & 0.2 & 2,11  \\

NGC\,3245 & U & 40.03 & 0.086 & 20.94 & 4 & 0.5 & 3,4 \\

NGC\,3489 & \nodata & 38.20 & 0.11 & $<$18.39 & 3 & \nodata & \nodata  \\ 

NGC\,3627 & S  & 39.16 & 0.13 & 19.15 & 2 & 1.2 & 2,12 \\

          & S+L &      &      & (20.45) & 2 & \nodata & \nodata \\

NGC\,3628 & S+D & 37.30 & 0.12 & 20.99 & 2 & 0.7 & 2,4 \\

NGC\,3675 & A & 38.22 & 0.12 & 19.70 & 2 & 1.0 & 2,4 \\

NGC\,3681 & \nodata & 38.29 & 0.14 & $<$19.84 & 2 &  \nodata & \nodata \\

NGC\,3692 & \nodata & \nodata & 0.12 & $<$20.73 & 5 & \nodata & \nodata \\

NGC\,3705 & \nodata & 39.15 & 0.079 & $<$20.24 & 5 & \nodata & \nodata \\

NGC\,3898 & S & 38.91 & 0.091 & 19.42 & 4 & \nodata & \nodata \\

NGC\,3917 & \nodata & 37.29 & $<$0.13 & $<$20.24 & 5 & \nodata & \nodata  \\

NGC\,3953 & \nodata & 38.79 & 0.12 & $<$20.28 & 5 & \nodata & \nodata \\

NGC\,3992 & \nodata & \nodata & 0.13 & $<$20.28 & 5 & \nodata & \nodata \\

NGC\,4013 & S+D & 37.82 & 0.11 & 20.45 & 2 & 0.6 & 2,4 \\

NGC\,4125 & \nodata & 39.09 & 0.14 & $<$19.55 & 3 & \nodata & \nodata \\

NGC\,4145 & \nodata & 37.75 & 0.13 & $<$20.45 & 5 & \nodata & \nodata \\

NGC\,4150 & \nodata & 38.18 & 0.13 & $<$18.75 & 3 & \nodata & \nodata  \\

NGC\,4192 & \nodata & 40.44 & 0.14 & $<$20.27 & 5 & \nodata & \nodata \\

NGC\,4216 & U & 38.89 & 0.12 & 19.50 & 6 & \nodata & \nodata \\

NGC\,4220 & \nodata & 39.40 & 0.12 & $<$20.31 & 5 & \nodata & \nodata \\

\enddata

\end{deluxetable}

\clearpage

\footnotesize
\begin{deluxetable}{lcrrrrrcrr}
\tablenum{5}
\tablecolumns{10} 
\tablewidth{0pc} 
\tablecaption{Summary of the complete composite galaxy sample. (cont.)}
\tablehead{ 
\colhead{} & \colhead{} & \colhead{Log L(H$\alpha$)} & \colhead{} & \colhead{log P$_{\rm 5\,GHz}$\tablenotemark{b}} & \colhead{} & \colhead {} & \colhead{}\\
\colhead{Galaxy} & \colhead{Morph.\tablenotemark{a}} & \colhead{(erg s$^{-1}$)} & \colhead{$\frac{[O~{\sc i}]}{H\alpha}$}  & \colhead{(W Hz$^{-1}$)} & \colhead{Ref.} & \colhead{$\alpha_{\rm radio}$} & \colhead{Ref.}\\
\colhead{(1)} & \colhead {(2)} & \colhead{(3)} & \colhead{(4)} & \colhead{(5)} & \colhead{(6)} & \colhead{(7)} & \colhead{(8)}}
\startdata 

NGC\,4281 & \nodata & 38.61 & $<$0.19 & $<$19.87 & 3 & \nodata & \nodata \\

NGC\,4321 & D &  39.37 & 0.11 & 21.02 & 2 & 1.3 & 7 \\

NGC\,4324 & \nodata & 38.95 & $<$0.16 & $<$19.87 & 3 & \nodata & \nodata  \\

NGC\,4350 & \nodata & \nodata & $<$0.18 & $<$19.23 & 3 & \nodata & \nodata \\

NGC\,4414 & D &  38.47 & $<$0.14 & 21.29 & 2 & 0.4 & 2,4  \\

NGC\,4419 & U & 40.72 & 0.075 & 20.07 & 6 & 0.7 & 4,6 \\

NGC\,4429 & \nodata & 39.18 & 0.097 & $<$20.19 & 5 & \nodata & \nodata \\

NGC\,4435 & \nodata & 39.67 & 0.13 & $<$20.19 & 5 & \nodata & \nodata  \\

NGC\,4459 & S & 38.79 & $<$0.13 & 19.08 & 4 & 0.7 & 3,4 \\

NGC\,4527 & S+D & 40.12 & 0.075 & 19.47 & 7 & 0.9 & 7 \\

NGC\,4552 & U & 38.55 & $<$0.13 & 20.78 & 2 & 0.2 & 2,4 \\

NGC\,4569 & \nodata & 40.28 & 0.062 & $<$20.19 & 5 & \nodata & \nodata \\

NGC\,4643 & A & 38.44 & 0.078 & 19.83 & 2 & \nodata & \nodata \\

NGC\,4713 & A & \nodata & 0.087 & 20.27 & 2 & \nodata & \nodata \\

NGC\,4826 & D & 38.87 & 0.073 & 21.20 & 2 & 0.3 & 2,4 \\

NGC\,5012 & A & 39.02 & 0.081 & 18.60 & 2 & \nodata & \nodata \\

NGC\,5055 & \nodata & 38.62 & $<$0.17 & $<$19.46 & 5 & \nodata & \nodata \\

NGC\,5354 & U & 38.71 & 0.14 & 21.02 & 2 & $-$0.2 & 2,4 \\

NGC\,5656 & A & 38.70 & $<$0.14 & 20.48 & 2 & \nodata & \nodata  \\

NGC\,5678 & D & 39.74 & 0.079 & 21.41 & 2 & \nodata & \nodata \\

NGC\,5701 & \nodata & 38.22 & $<$0.14 & $<$20.58 & 5 & \nodata  & \nodata \\

NGC\,5746 & S & 39.25 & 0.10 & 19.38 & 4 & \nodata & \nodata \\

NGC\,5866 & S & 38.82 & $<$0.13 & 20.37 & 8 & 0.6 & 3,4 \\

NGC\,5838 & U & \nodata & 0.053 & 20.18 & 2 & 0.0 & 1,4 \\

NGC\,5846 & U & 39.01 & 0.091 & 20.68 & 2 & 0.4 & 2,4  \\

NGC\,5879 & \nodata & 38.86 & 0.16 & $<$19.52 & 2 & \nodata & \nodata \\

NGC\,5921 & S & 39.15 & 0.11 & 20.08 & 2 & 0.6 & 2,4 \\

NGC\,6384 & \nodata & 38.12 & $<$0.15 & $<$19.92 & 2 & \nodata & \nodata \\

NGC\,6482 & \nodata & 39.23 & $<$0.13 & $<$20.51 & 2 & \nodata & \nodata  \\

NGC\,6503 & \nodata & 37.56 & 0.080 & $<$18.64 & 2 & \nodata & \nodata \\

NGC\,7177 & S & 39.38 & 0.14 & 19.81 & 1 & \nodata & \nodata \\

NGC\,7331 & U & 38.73 & $<$0.097 & 18.48 & 9 & 0.5 & 9 \\

          & D &       &          & (21.37) & 1 & \nodata    & \nodata   \\

NGC\,7742 & D & 39.07 & 0.13 & 21.39 & 1 & \nodata & \nodata \\

\enddata

\end{deluxetable}

\clearpage

\footnotesize
\begin{deluxetable}{lcrrrrrcrr}
\tablenum{5}
\tablecolumns{10} 
\tablewidth{0pc} 
\tablecaption{Summary of the H~{\sc ii} galaxies (cont.).}
\tablehead{ 
\colhead{} & \colhead{} & \colhead{Log (H$\alpha$)} & \colhead{} & \colhead{log P$_{\rm 5\,GHz}$\tablenotemark{b}} & \colhead{} & \colhead {} & \colhead {} \\
\colhead{Galaxy} & \colhead{Morph.\tablenotemark{a}}  & \colhead{(erg s$^{-1}$)} & \colhead{$\frac{[O~{\sc i}]}{H\alpha}$} & \colhead{(W Hz$^{-1}$)} & \colhead{Ref.} & \colhead{$\alpha_{\rm radio}$} & \colhead{Ref.}\\
\colhead{(1)} & \colhead {(2)} & \colhead{(3)} & \colhead{(4)} & \colhead{(5)} & \colhead{(6)} & \colhead{(7)} & \colhead{(8)}}
\startdata 

NGC\,3593 & D & 38.90 & 0.042 & 19.93 & 2 & 0.3 & 2,4 \\ 

NGC\,3684 & \nodata & 38.95 & 0.030 & $<$19.59 & 2 & \nodata & \nodata  \\

NGC\,4100 & D & 39.95 & 0.012 & 20.47 & 2 & 0.5 & 2,4 \\

NGC\,4217 & D & 38.62 & 0.049 & 20.95 & 2 & 0.0 & 2,4 \\

NGC\,4245 & \nodata & 37.75 & $<$0.038 & $<$18.82 & 2 & \nodata & \nodata \\

NGC\,4369 & D & 39.76 & 0.014 & 20.40 & 2 & 0.1 & 2,4 \\

NGC\,4405 & \nodata & 39.64 & 0.026 & $<$19.85 & 2 & \nodata & \nodata \\

NGC\,4424 & D & 39.56 & 0.014 & 20.00 & 2 & $-$0.2 & 2,4 \\

NGC\,4470 & A & \nodata & 0.023 & 20.56 & 2 & \nodata & \nodata \\

NGC\,4710 & D & 39.02 & 0.045 & 20.38 & 2 & 0.5 & 2,4 \\

NGC\,4800 & A & 39.93 & 0.041 & 19.59 & 2 & 1.3 & 2,4 \\

NGC\,4845 & D & 39.90 & 0.034 & 20.63 & 2 & 0.5 & 2,4 \\

\enddata

\tablenotetext{a}{U  -  single  unresolved  radio  core;  S  -  single
slightly resolved radio  core; D - diffuse radio  emission; L - linear
or multicomponent structure; A - ambiguous.}

\tablenotetext{b}{Extrapolation  to 5\,GHz was  made using  a spectral
index  of  0.7  for  all  sources and  integrated  flux  densities  of
NGC\,660, NGC\,1055, NGC\,3627, and NGC\,7331.  Integrated power is in
parentheses.}

\tablerefs{(1) - This paper,  8.4\,GHz 2\arcsecpoint5; (2) Filho \etal
2000, 8.4\,GHz,  2\arcsecpoint5; (3) Wrobel \&  Heeschen 1991, 5\,GHz,
5\arcsec; (4) Becker \etal 1995 (FIRST), 1.4\,GHz, 5\arcsec; (5) Nagar
\etal 2000,  15\,GHz, 0\arcsecpoint2;  (6) Nagar \etal  2000, 15\,GHz,
4\arcsec; (7)  Vila \etal  1990; 1.4\,GHz, 1\arcsecpoint4  and 5\,GHz,
1\arcsecpoint2; (8) Falcke  \etal 2000, 5\,GHz, 0\arcsecpoint0025; (9)
Cowan \etal 1994, 1.4\,GHz, 1\arcsecpoint4 and 5\,GHz, 1\arcsecpoint2;
(10) Downes  \etal 1986,  5\,GHz, 7\arcsec;  (11) Condon  \etal 1998b,
1.4\,GHz, 5\arcsec; (12) Saikia \etal 1994, 5\,GHz, 1\arcsecpoint2.}

\end{deluxetable}

\clearpage

\clearpage

\begin{figure}
\plottwo{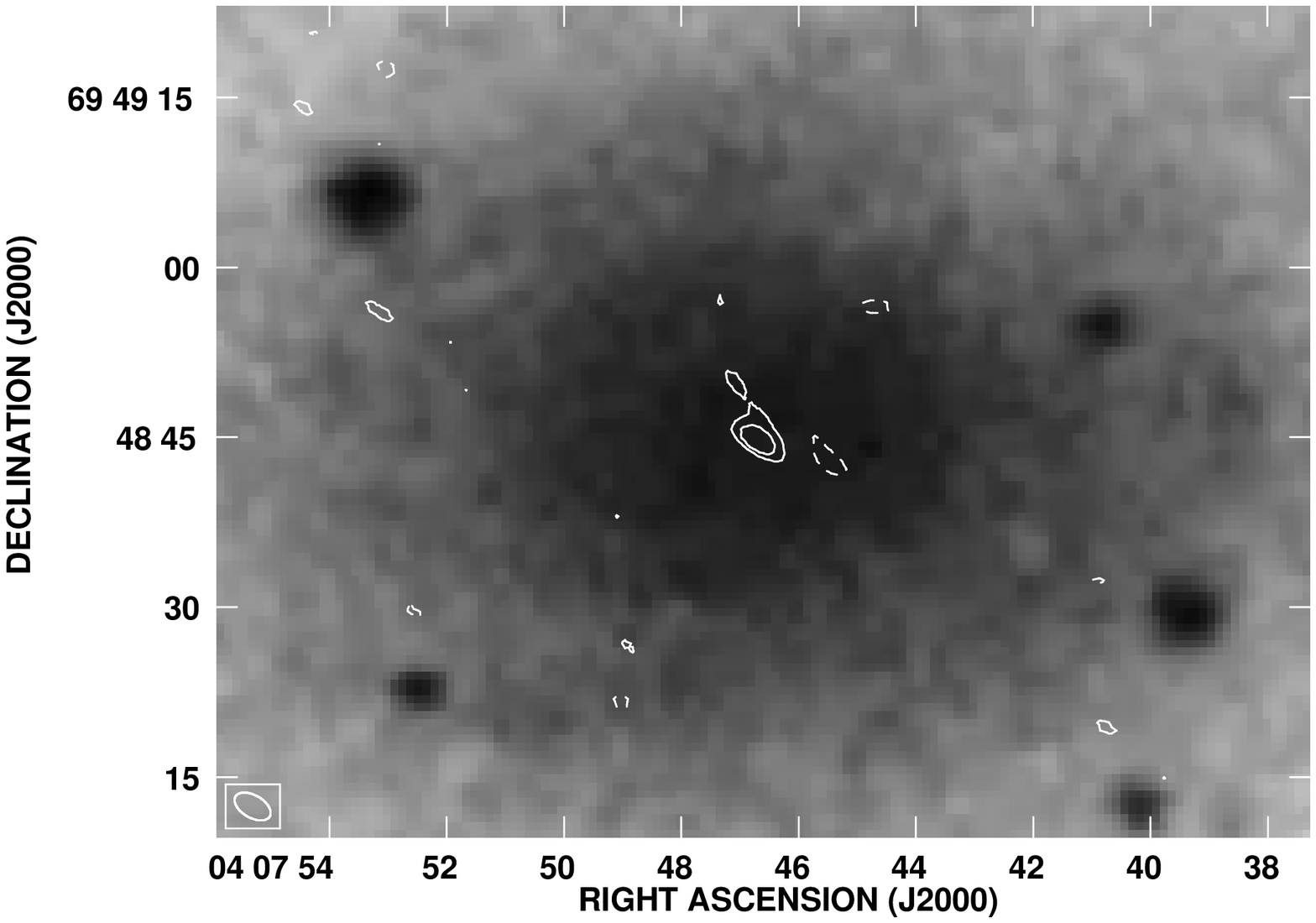}{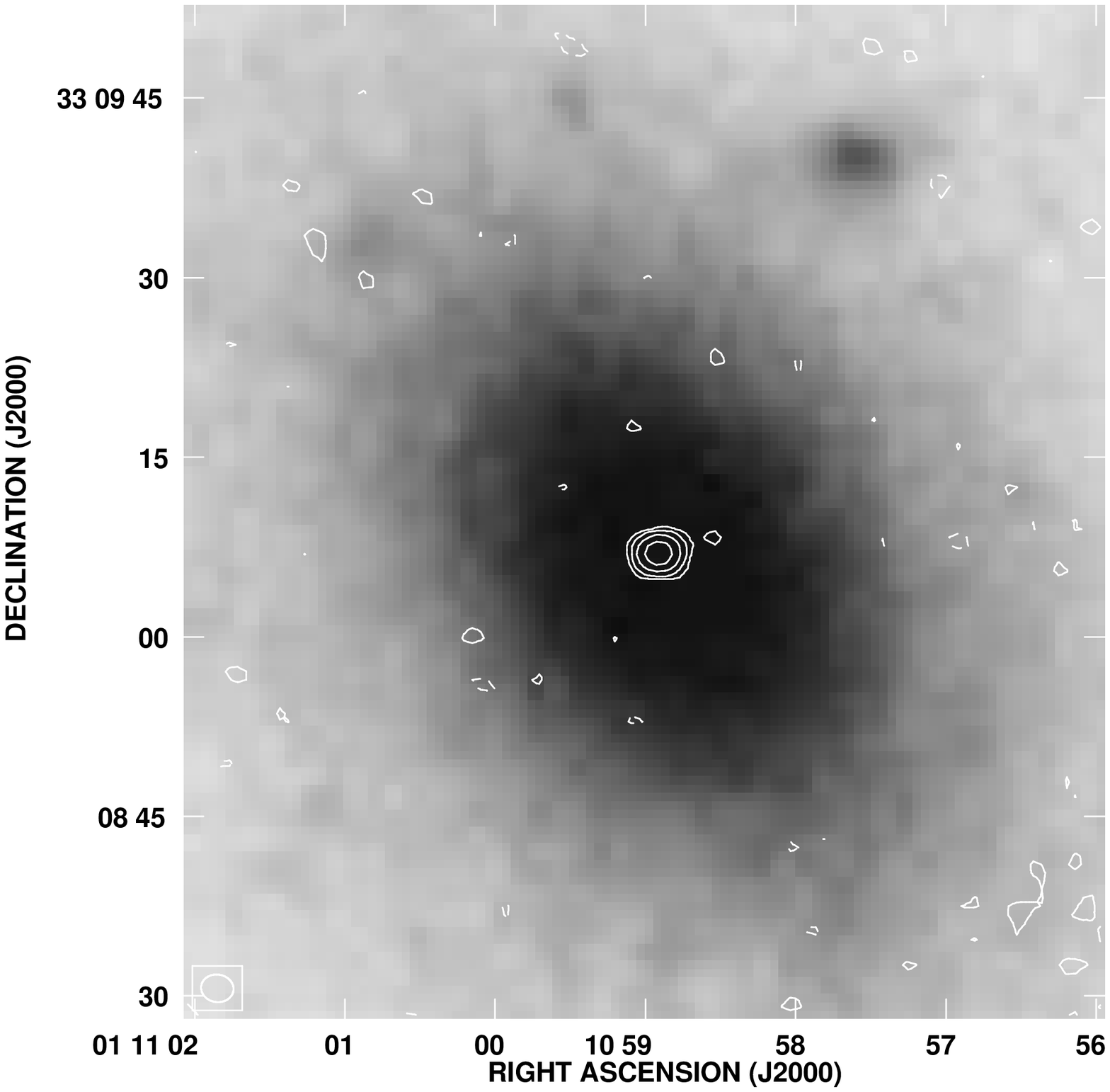}
\plottwo{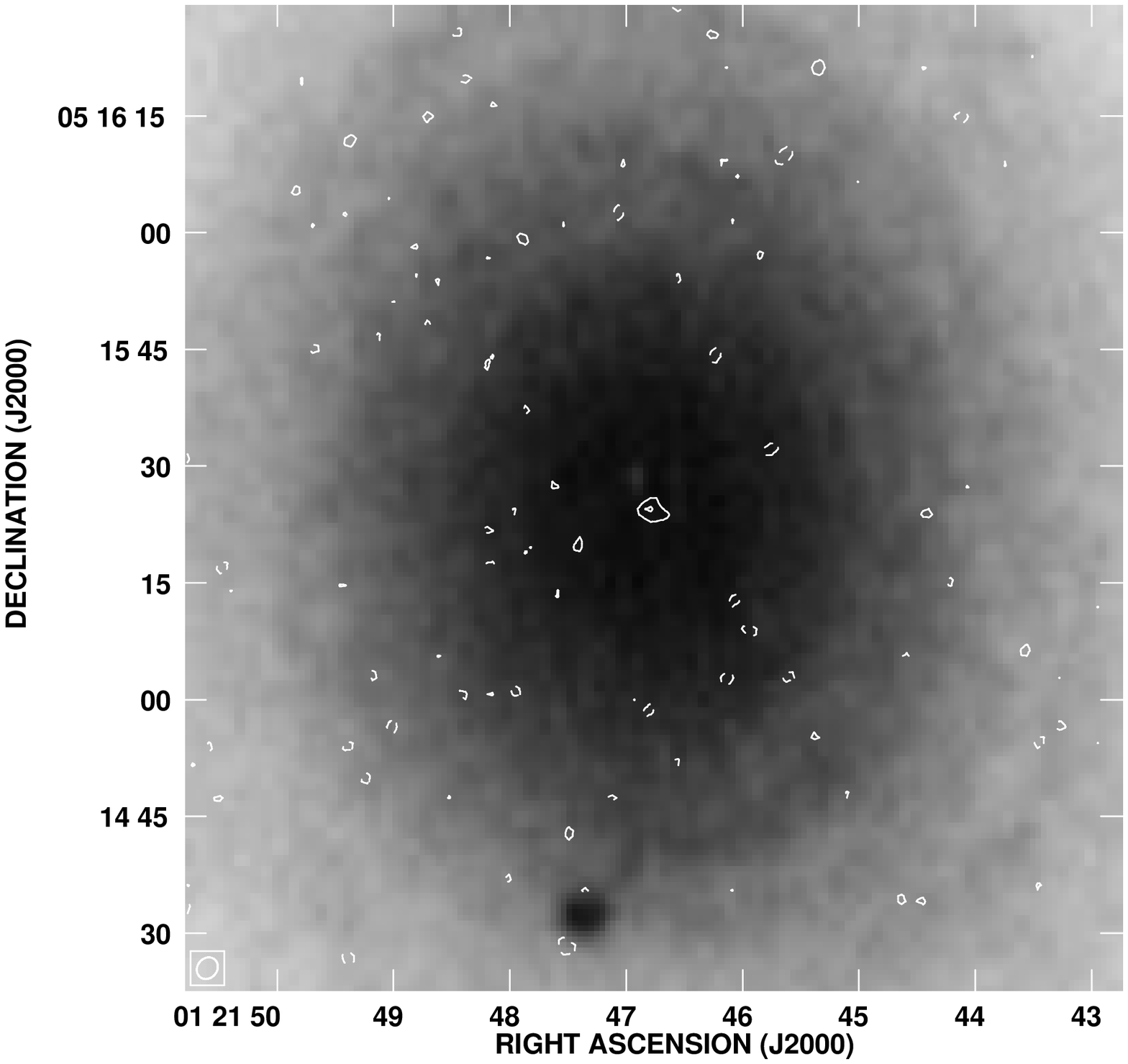}{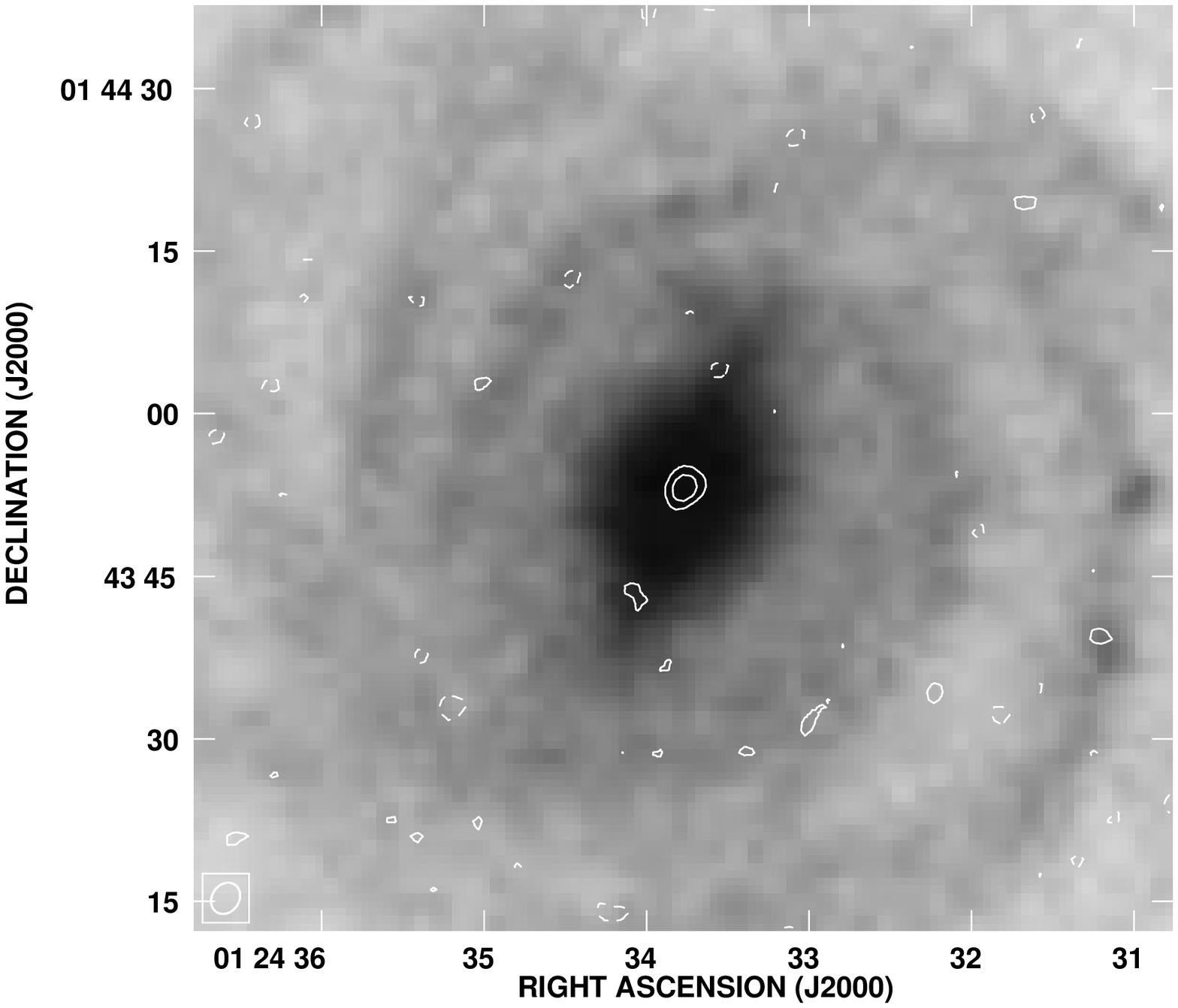}
\caption{Radio emission (contours) superimposed on optical images from
the  Digitized  Sky  Survey  (greyscale).   Contour  levels  are  CLEV
$\times$ (--3,3,6,12,24,48,96), where CLEV is the rms noise level (see
Table~2). The size of the restoring beam is given in parentheses after
each object name (see Table~2).  For NGC\,660, NGC\,1055 and NGC\,7331
we present greyscale  radio images with the scale  levels given on the
top  of  each  map.   {\bf  (1a)}  IC\,356  (3\arcsecpoint50  $\times$
1\arcsecpoint98),   {\bf  (1b)}  NGC\,410   (2\arcsecpoint73  $\times$
2\arcsecpoint30),   {\bf  (1c)}  NGC\,488   (2\arcsecpoint93  $\times$
2\arcsecpoint52)  and {\bf  (1d)}  NGC\,521 (3\arcsecpoint04  $\times$
2\arcsecpoint50).}

\end{figure}

\clearpage

\begin{figure}
\plottwo{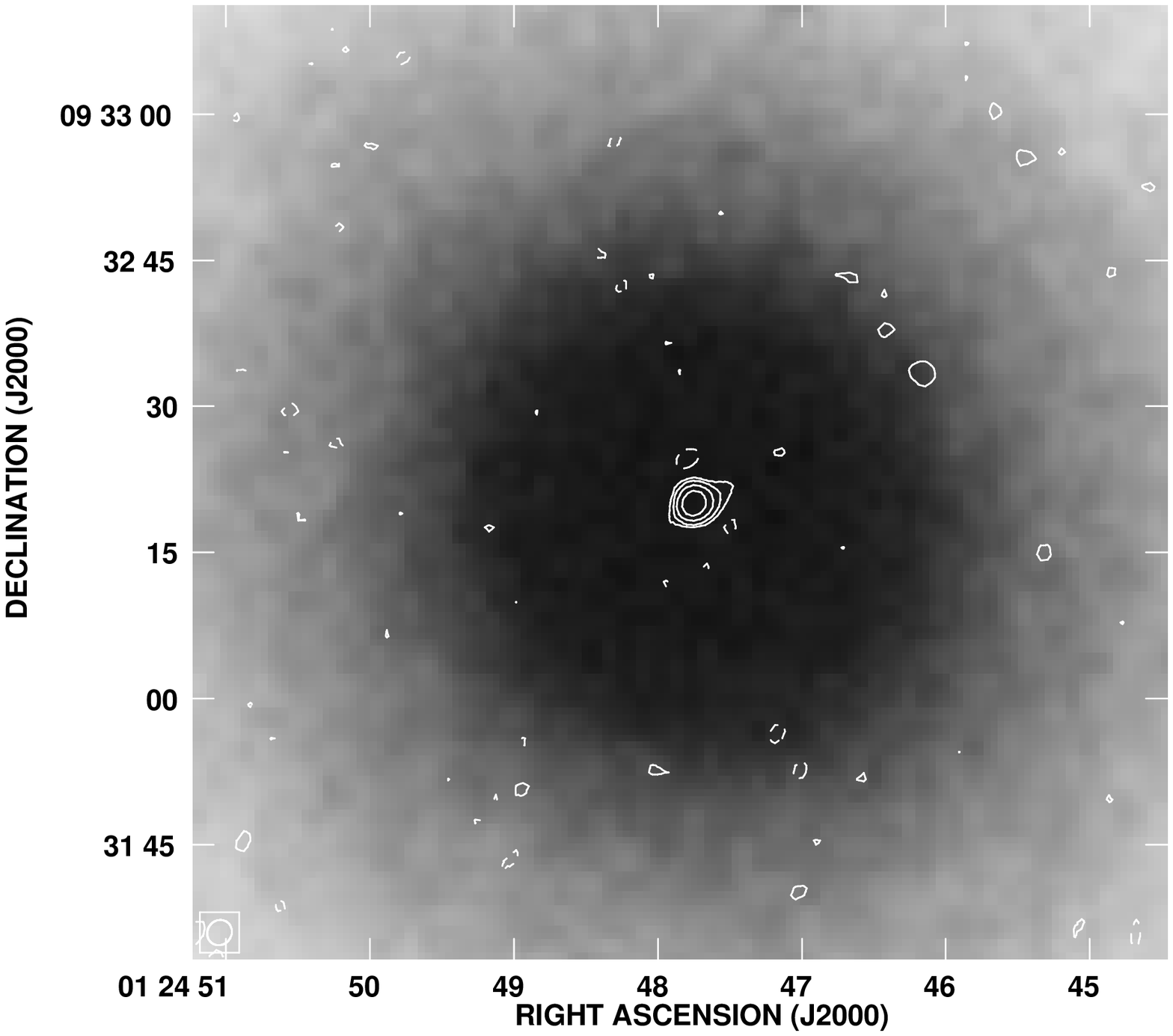}{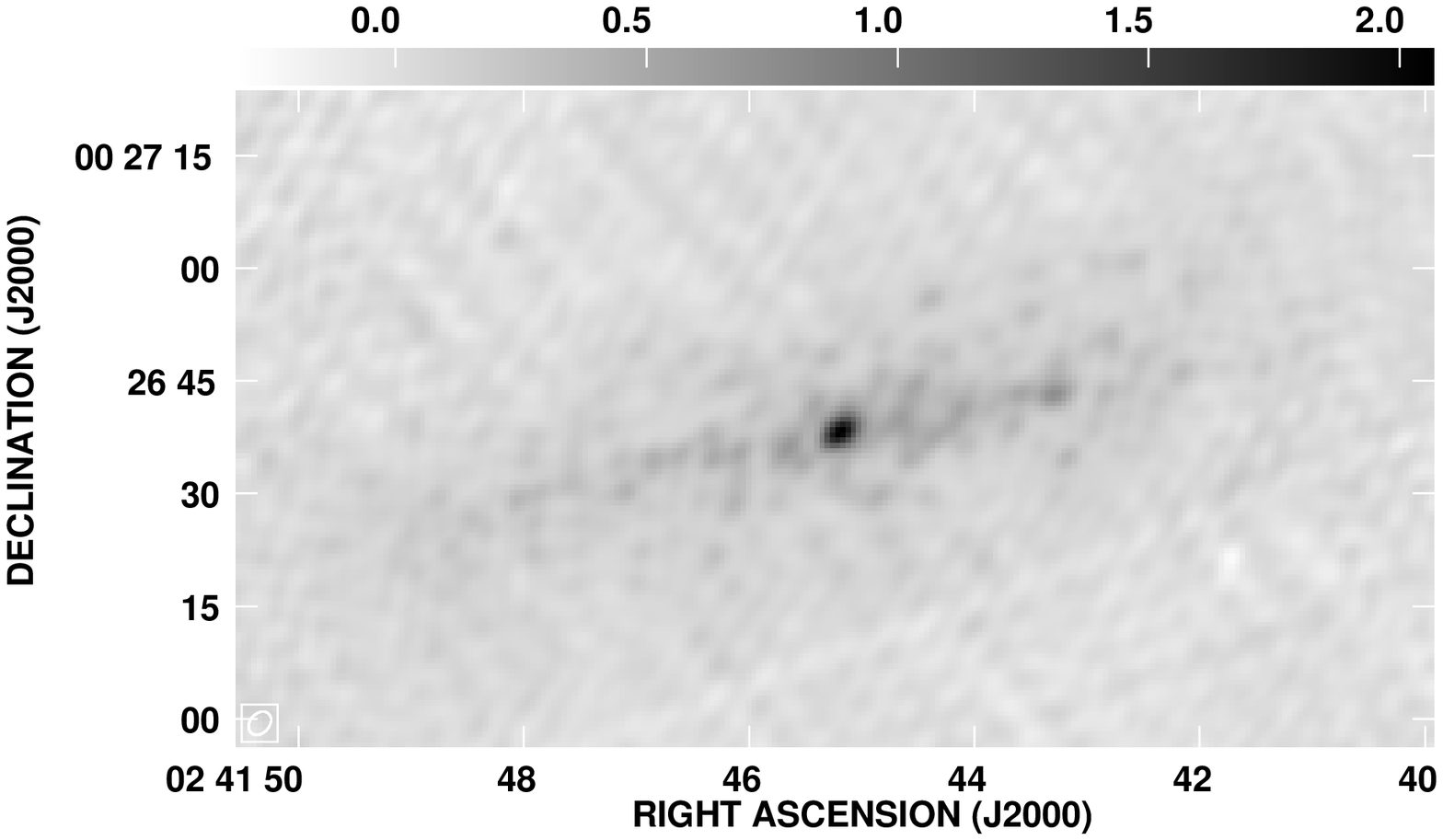}
\plottwo{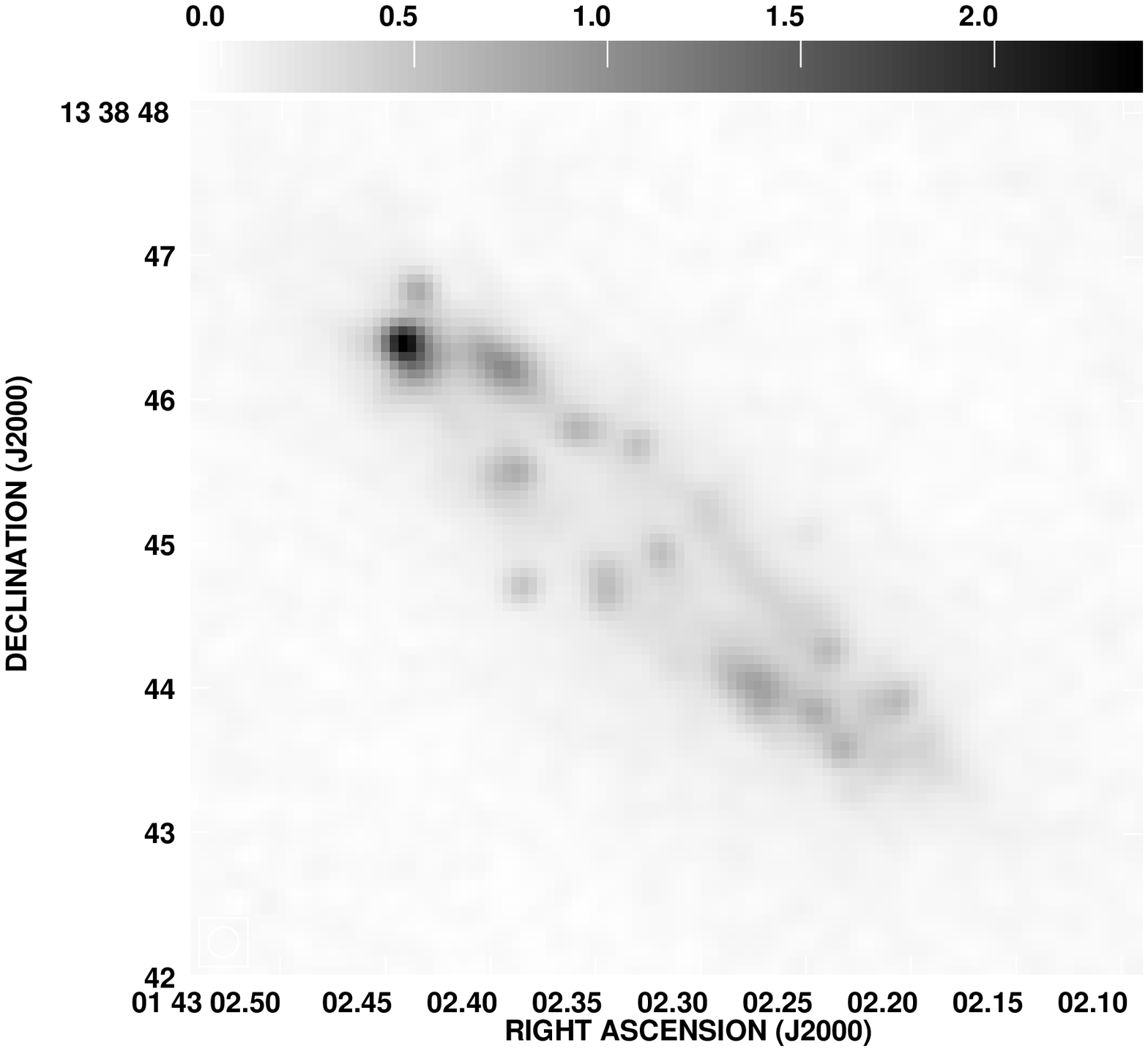}{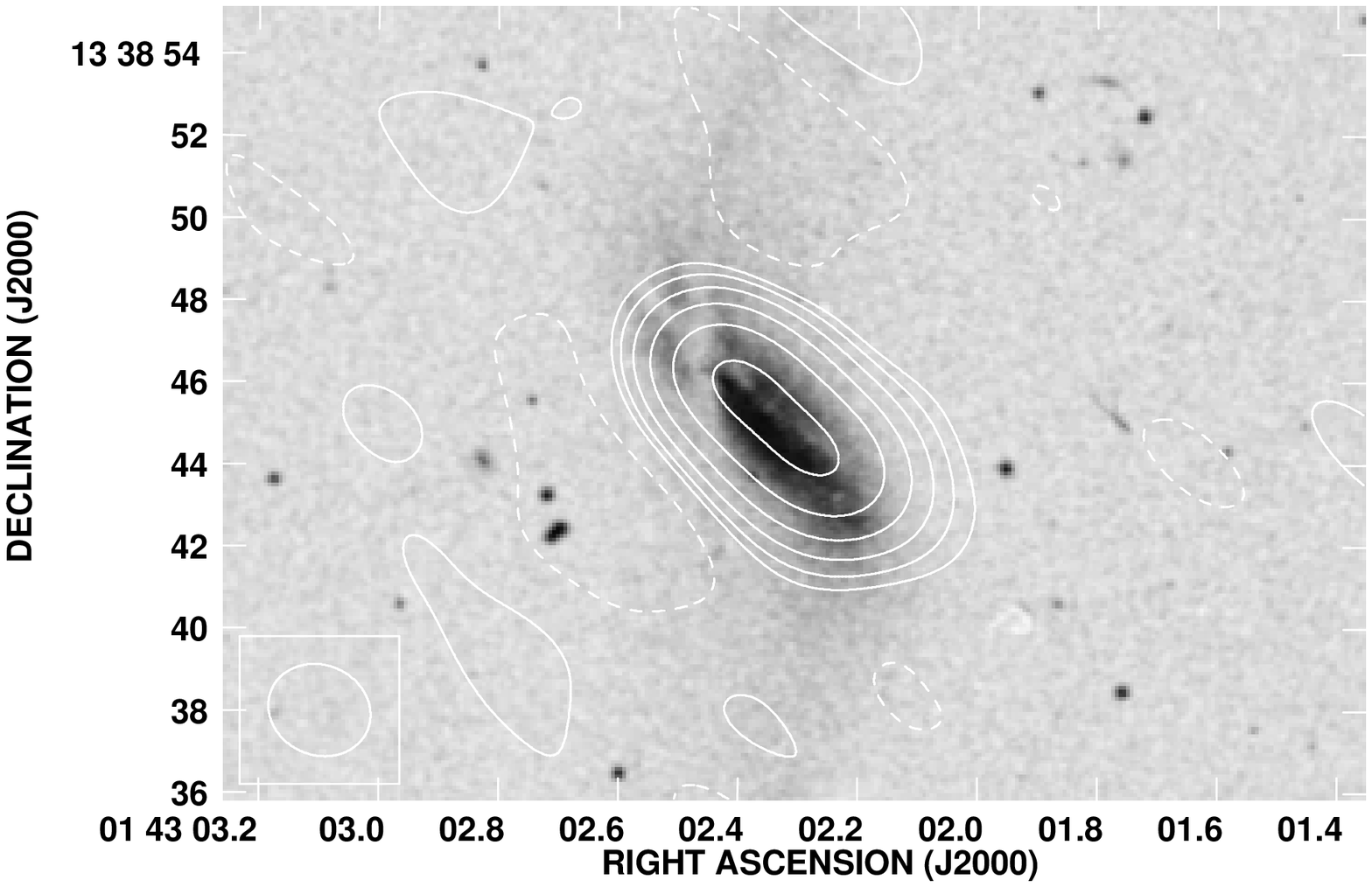}
\caption{As  in   Figure~1.   {\bf  (2a)}   NGC\,524  (2\arcsecpoint66
$\times$  2\arcsecpoint47),   {\bf  (2b)}  NGC\,1055  (3\arcsecpoint45
$\times$   2\arcsecpoint57)  {\bf   (2c)}   NGC\,660  (0\arcsecpoint21
$\times$  0\arcsecpoint21), and  {\bf (2d)}  NGC\,660 (2\arcsecpoint54
$\times$ 2\arcsecpoint19).}

\end{figure}

\clearpage

\begin{figure}
\plottwo{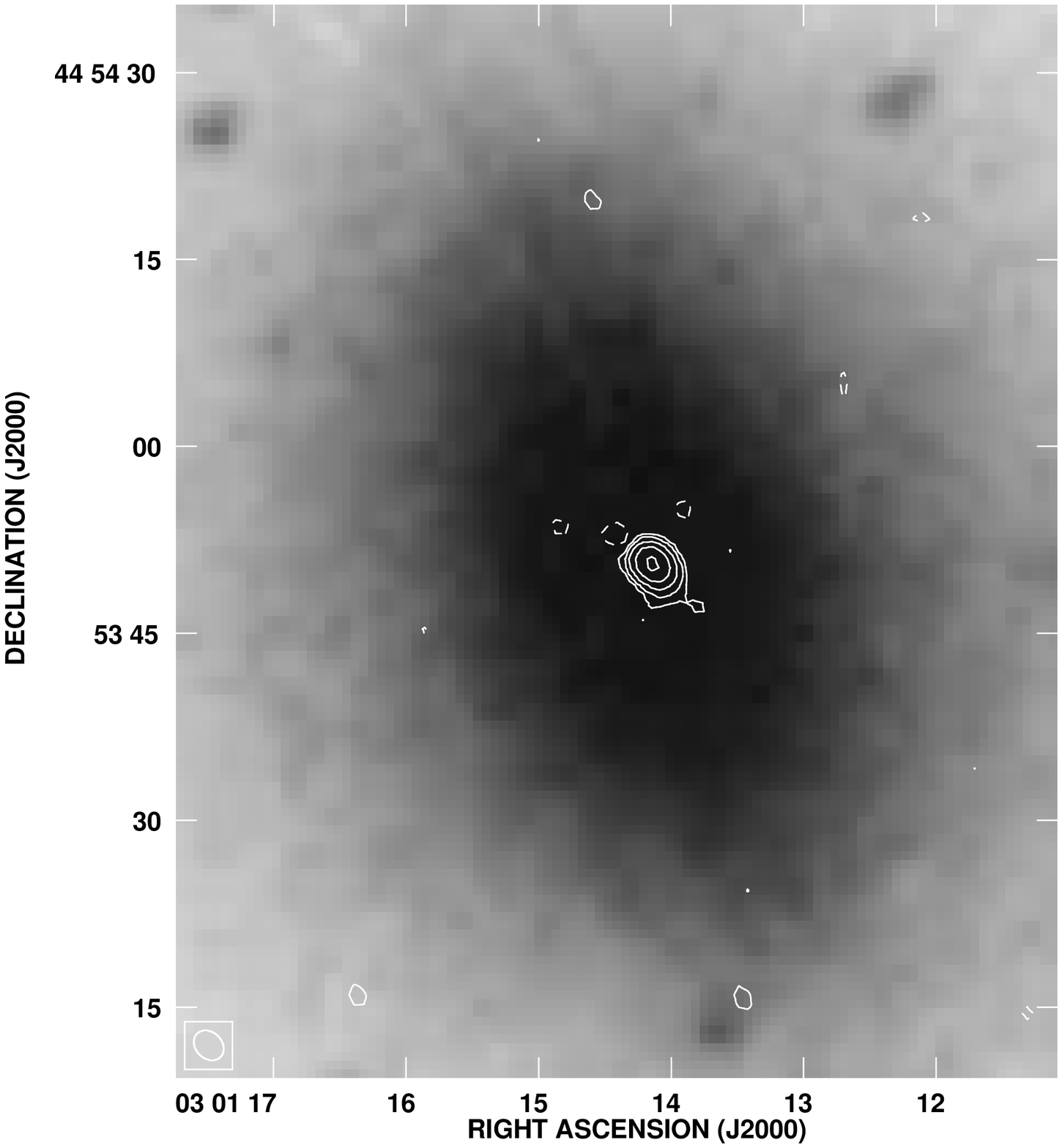}{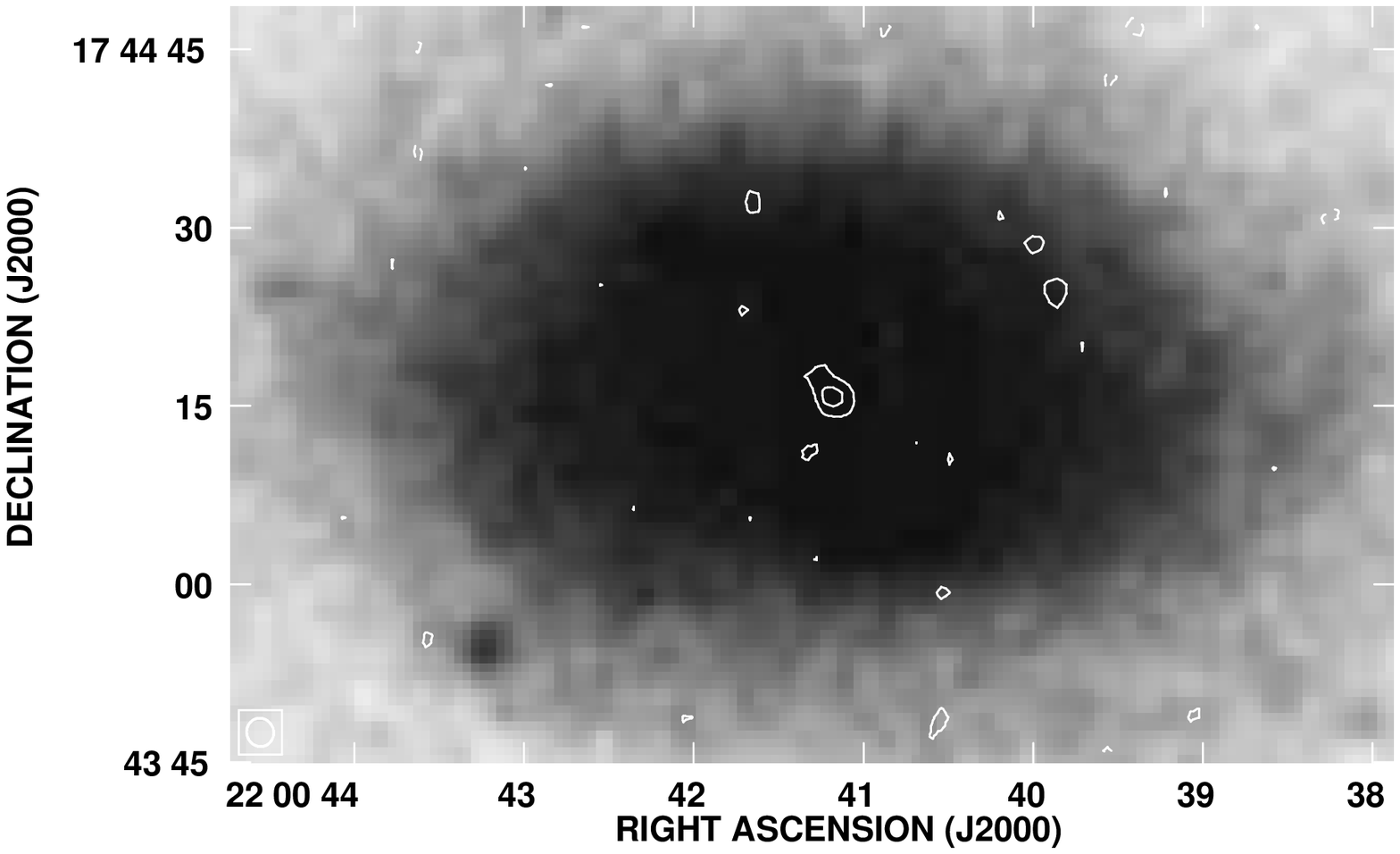}
\plottwo{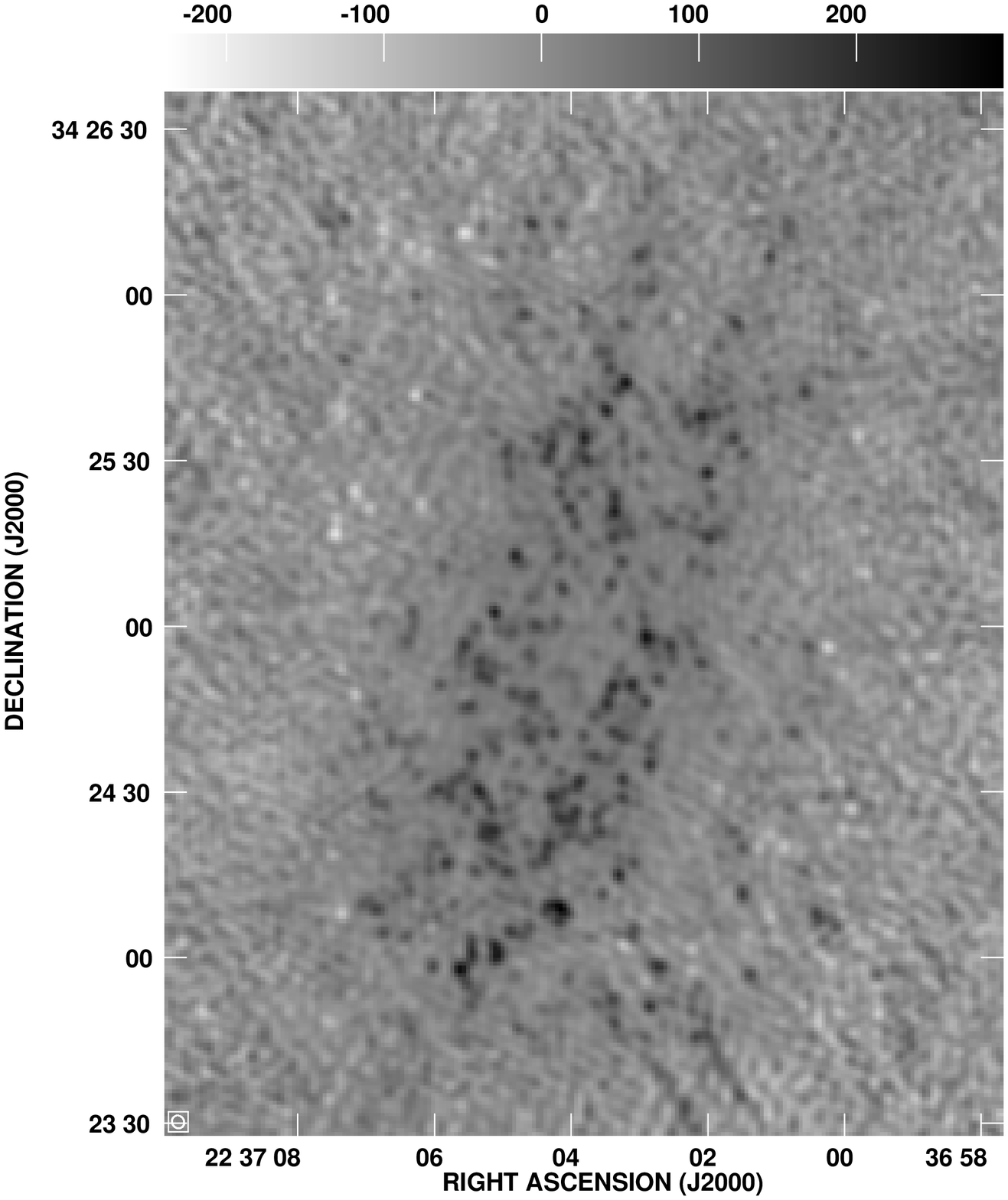}{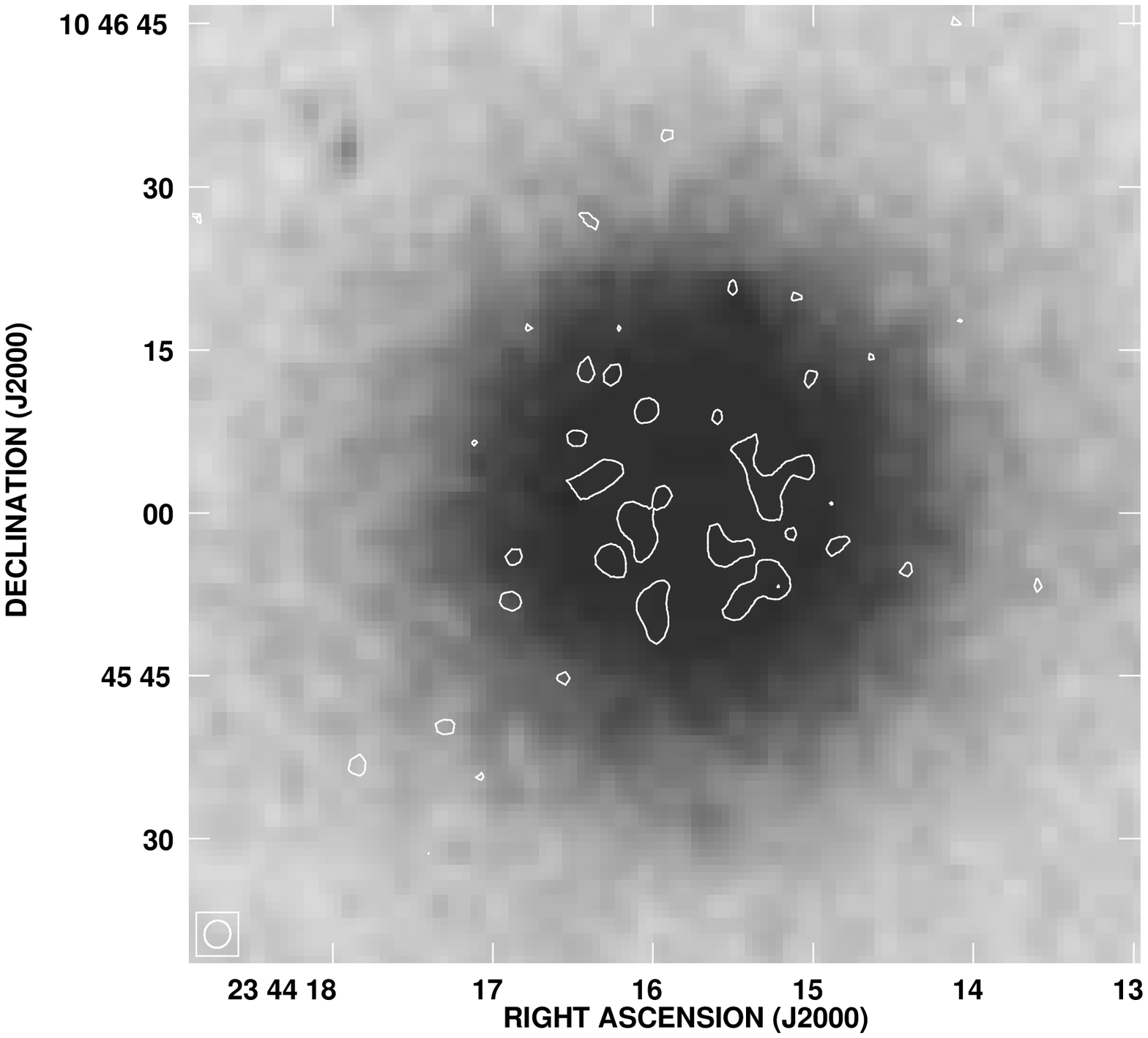}
\caption{As  in  Figure~1.    {\bf  (3a)}  NGC\,1161  (2\arcsecpoint66
$\times$  2\arcsecpoint10),   {\bf  (3b)}  NGC\,7177  (2\arcsecpoint40
$\times$  2\arcsecpoint26),   {\bf  (3c)}  NGC\,7331  (2\arcsecpoint37
$\times$ 2\arcsecpoint15),  and {\bf (3d)}  NGC\,7742 (2\arcsecpoint52
$\times$ 2\arcsecpoint38).}

\end{figure} 

\clearpage

\begin{figure}
\epsscale{.3}
\plottwo{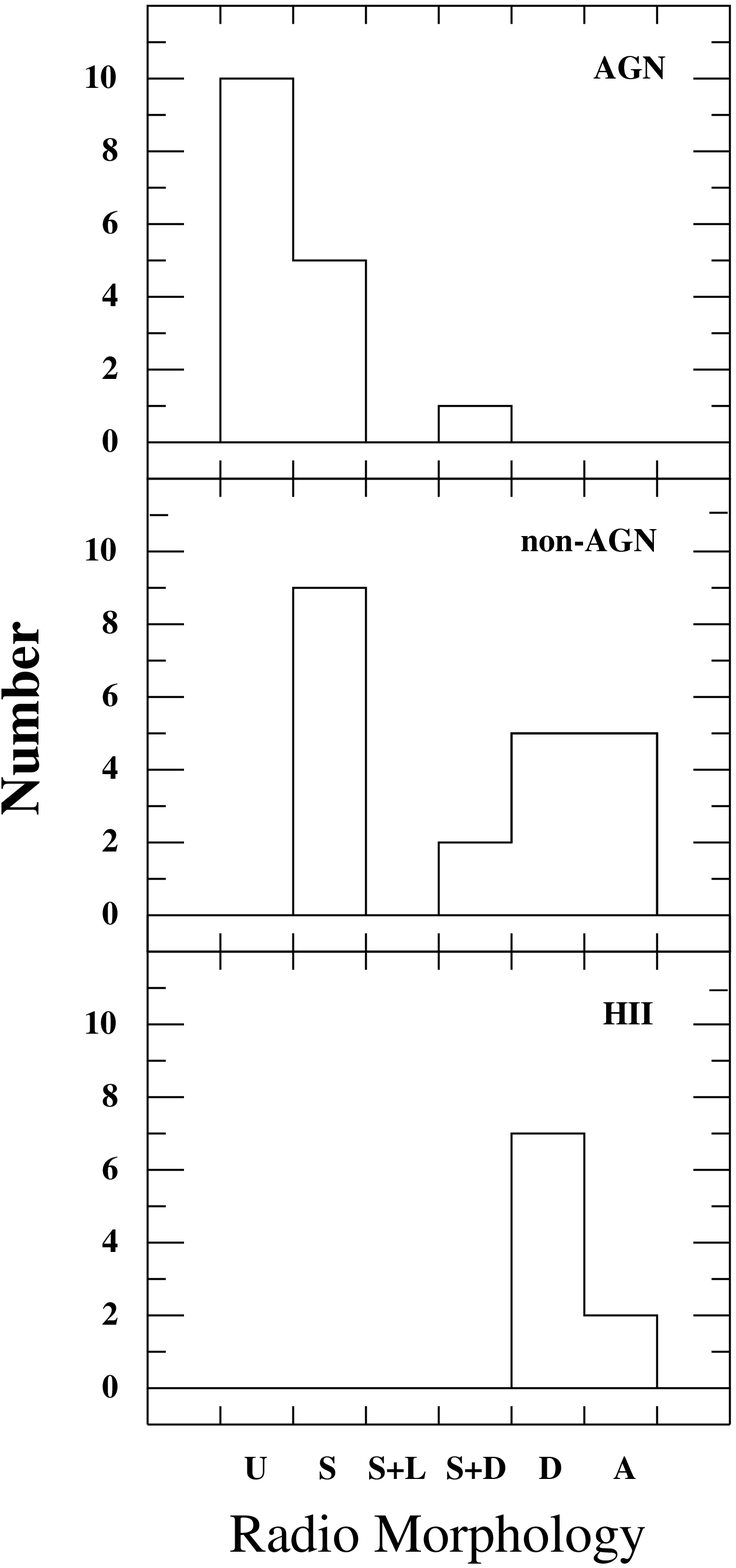}{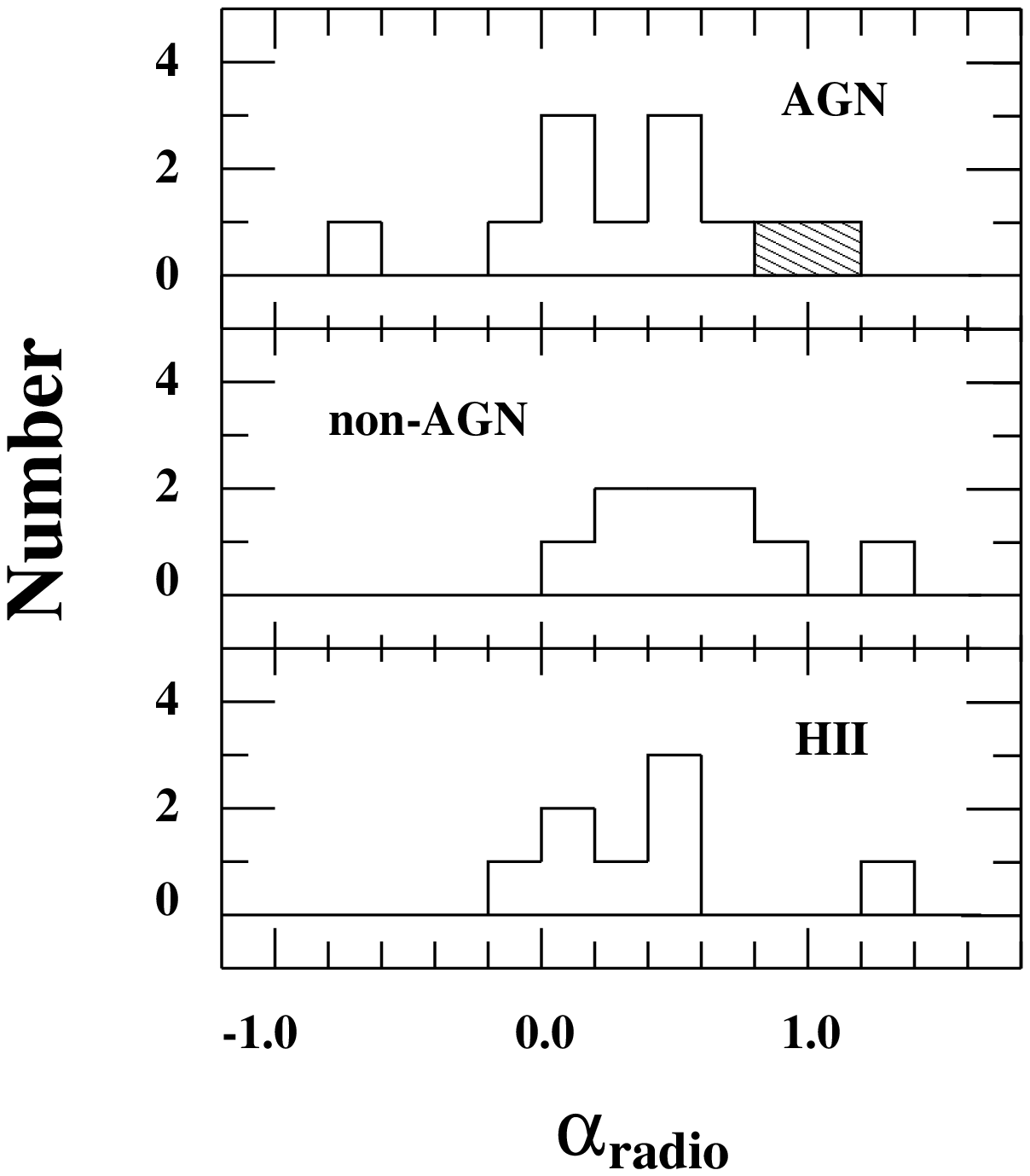}
\end{figure}

\clearpage

\begin{figure}
\plottwo{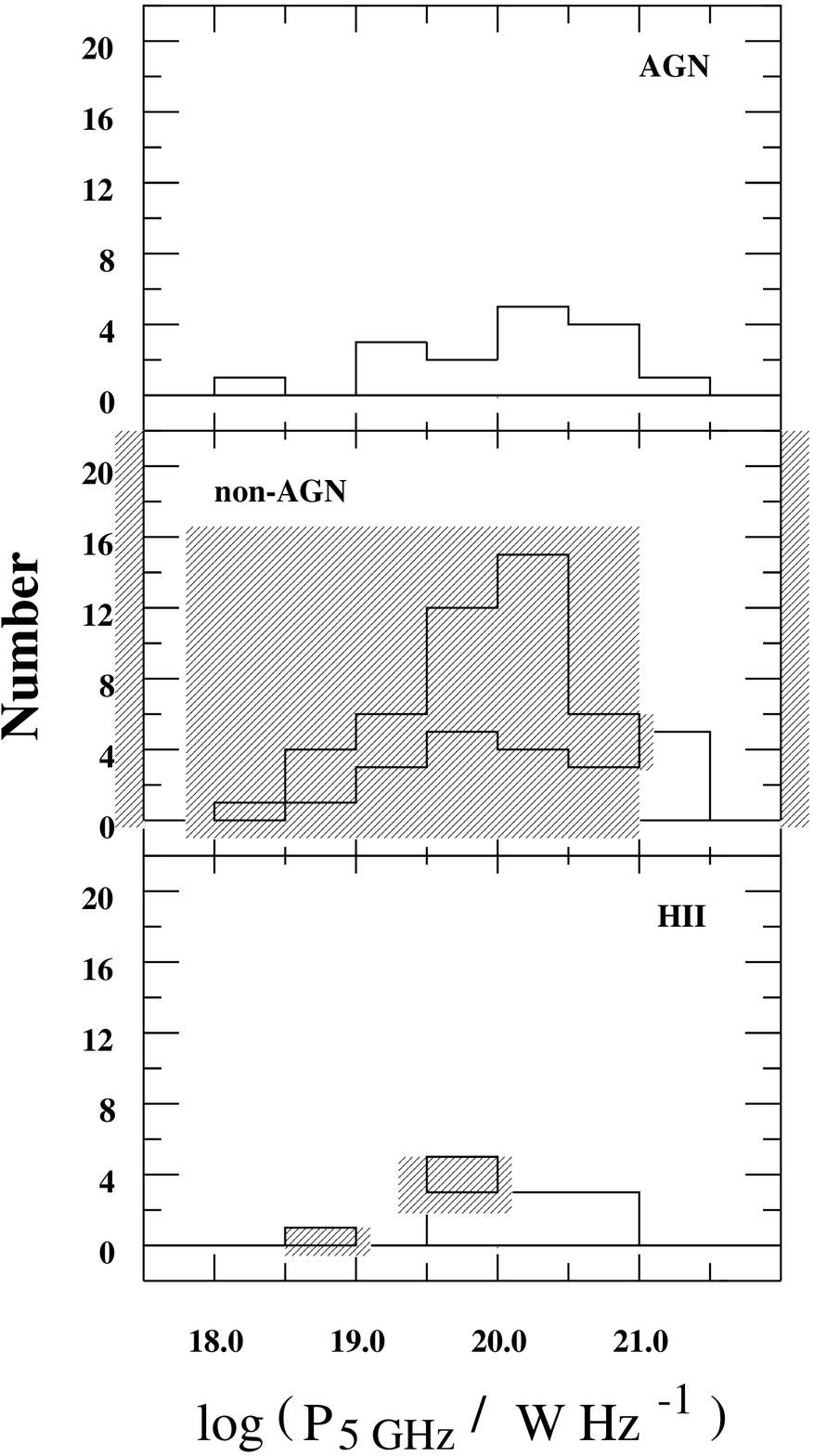}{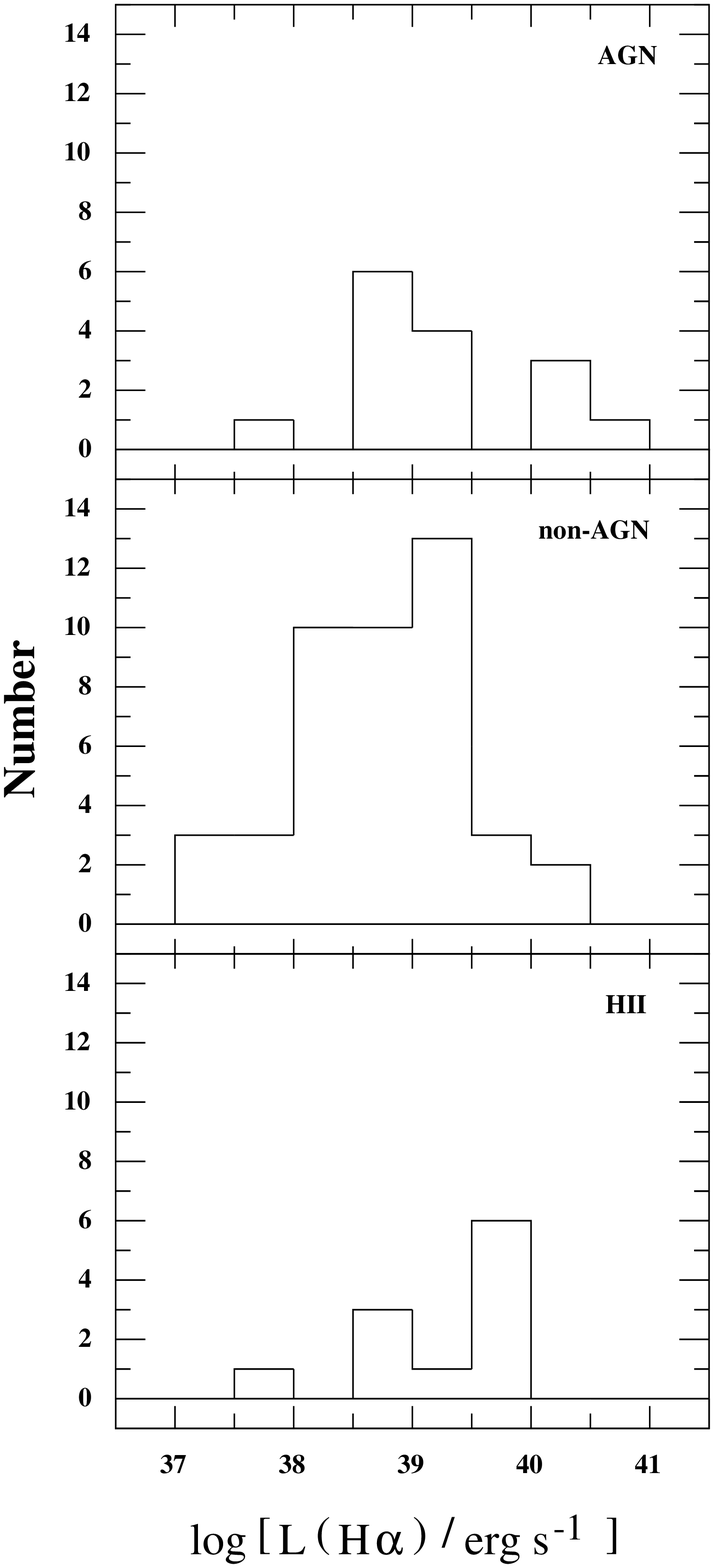}
\caption{Histograms  of  the  nuclear  radio  morphology  (4a),  radio
spectral  index  (4b),  5\,GHz  power  (4c)  and  extinction-corrected
H$\alpha$  luminosity   (4d).   Morphological  classification   is  as
follows:  U  - single  unresolved  radio  core;  S -  single  slightly
resolved  radio  core; D  -  diffuse radio  emission;  L  - linear  or
multicomponent structure; A -  ambiguous.  Upper limits to radio power
are indicated  by hatched regions.  For the  spectral indices, hatched
regions refer  to AGN-type sources  whose radio core flux  density may
have been contaminated by radio emission from the host galaxy.}

\end{figure}

\clearpage

\begin{figure}
\epsscale{1.0}
\plotone{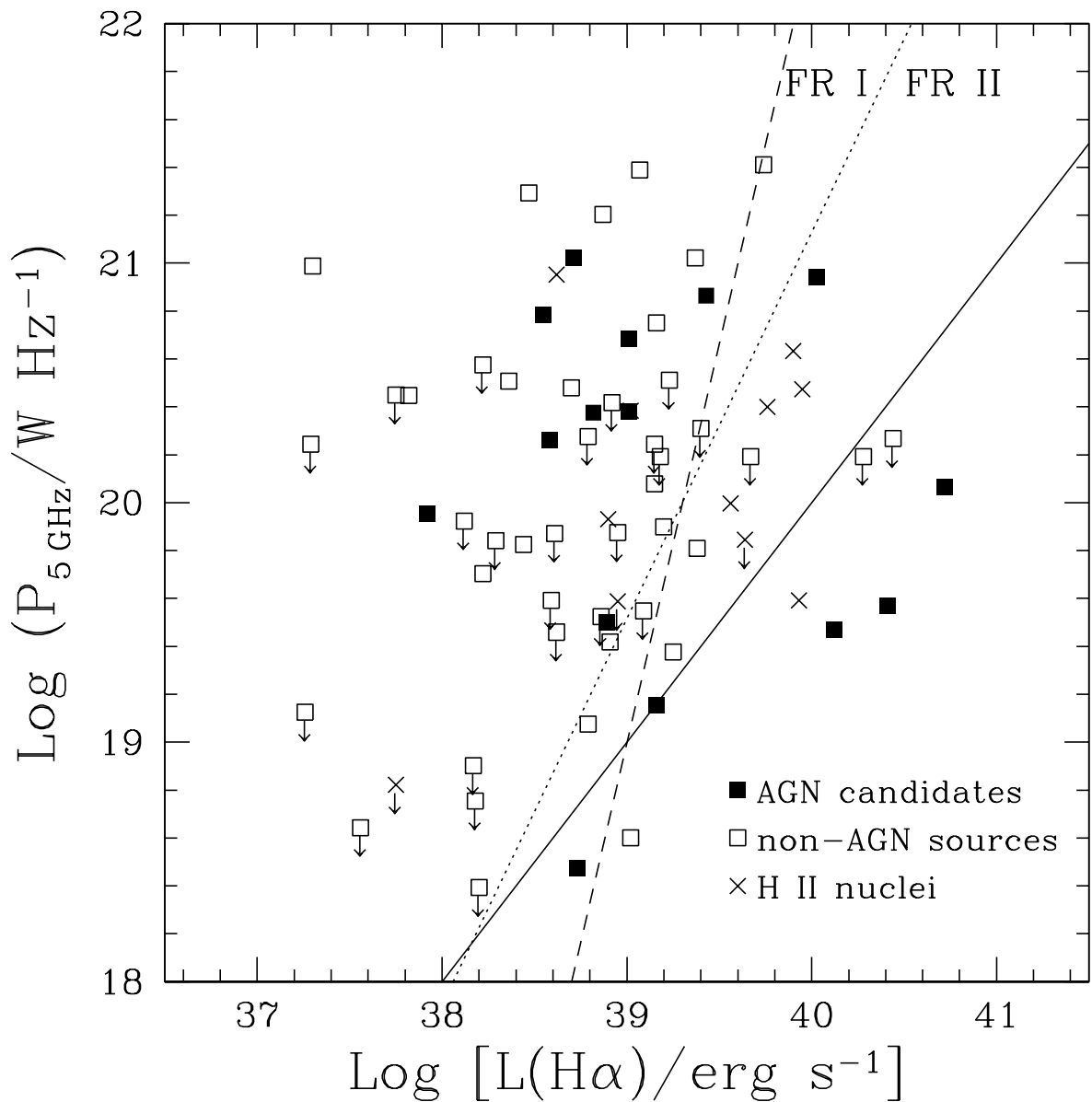}
\caption{Plot of the  nuclear 5\,GHz power versus extinction-corrected
nuclear H$\alpha$  luminosity.  Upper  limits to P$_{\rm  5\,GHz}$ are
designated by arrows.  The solid  line defines the thermal radio power
predicted from  the H$\alpha$ line luminosity.  The  dashed and dotted
lines are, respectively, the  best-fitting linear correlations for the
radio cores in FR~I and FR~II radio galaxies.}

\end{figure}

\clearpage
\begin{figure}
\plotone{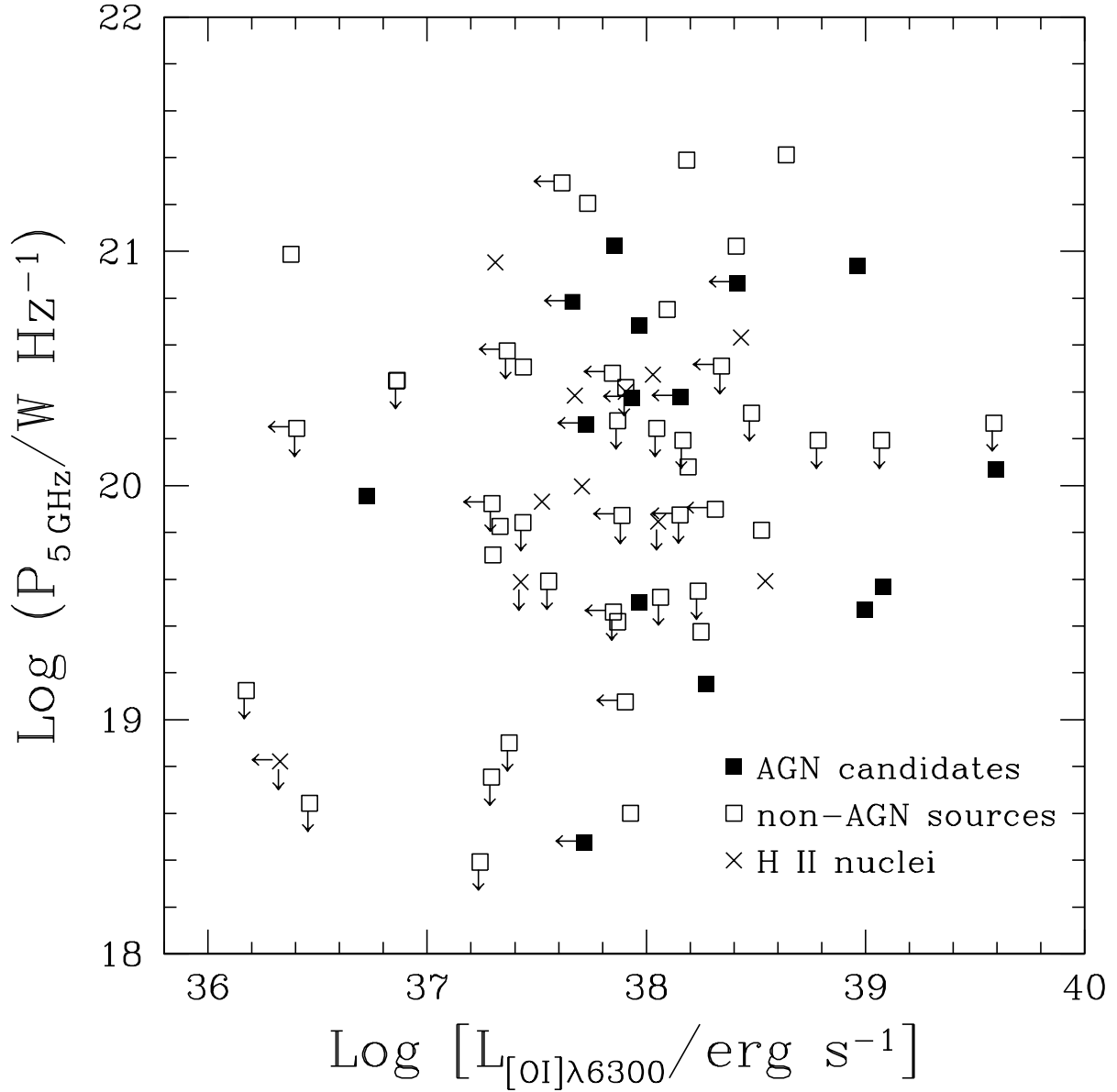}
\caption{Plot of the  nuclear 5\,GHz power versus extinction-corrected
[O~{\sc i}]$\lambda$6300 luminosity.  Upper limits to P$_{\rm 5\,GHz}$
and [O~{\sc i}] line luminosity are designated by arrows.}

\end{figure}

\end{document}